\documentclass[12pt,preprint]{aastex}

\slugcomment{Accepted for publication in ApJ}

\shorttitle{Molecular Emission from cB58}
\shortauthors{Baker et al.}

\received{2003 September 6}
\begin{document}

\title{Molecular Gas in the Lensed Lyman Break Galaxy cB58\altaffilmark{1}}

\altaffiltext{1}{Based on observations carried out with the IRAM Plateau de 
Bure Interferometer.  IRAM is supported by the INSU/CNRS (France), the 
MPG (Germany), and the IGN (Spain).}

\author{Andrew J. Baker, Linda J. Tacconi, Reinhard Genzel\altaffilmark{2}, 
Matthew D. Lehnert, \& Dieter Lutz}

\affil{Max--Planck--Institut f{\" u}r extraterrestrische Physik,
Postfach 1312, D--85741 Garching, Germany}

\altaffiltext{2}{Also: Department of Physics, 366 Le~Conte Hall, University of 
California, Berkeley, CA 94720--7300}

\email{\{ajb,linda,genzel,mlehnert,lutz\}@mpe.mpg.de}

\begin{abstract}
We have used the IRAM Plateau de Bure Interferometer to map CO(3--2) emission 
from the gravitationally lensed Lyman break galaxy MS\,1512--cB58.  This is 
the first detection of a molecular emission line in any Lyman break system; 
its integrated intensity implies a total molecular gas mass of 
$6.6^{+5.8}_{-4.3} \times 10^9\,h_{0.7}^{-2}\,M_\odot$, while its width 
implies a dynamical mass of $1.0^{+0.6}_{-0.4} \times 10^{10}\,{\rm csc}^2 
i\,h_{0.7}^{-1}\,M_\odot$ (for a flat $\Omega_\Lambda = 0.7$ 
cosmology).  These estimates are in excellent concordance with nearly all 
parameters of the system measured at other wavelengths, and yield a consistent 
picture of past and future star formation with no obvious discrepancies 
requiring explanation by differential lensing.  In particular, we find that 
the age and remaining lifetime of the current episode of star formation are 
likely to be similar; the surface densities of star formation and molecular 
gas mass are related by a Schmidt law; and the fraction of baryonic mass 
already converted into stars is sufficient to account for the observed 
enrichment of the interstellar medium to $0.4\,Z_\odot$.  Barring substantial 
gas inflow or a major merger, the stars forming in the current episode 
will have mass and coevality at $z = 0$ similar to those of a spiral bulge.  
Assuming cB58 is a typical Lyman break galaxy apart from its magnification, 
its global parameters suggest that the prescriptions for star formation used 
in some semi--analytic models of galaxy evolution require moderate 
revision, although the general prediction that gas mass fraction should 
increase with redshift is validated.  The length of cB58's star formation 
episode relative to the time elapsed over the redshift range $2.5 \leq z \leq 
3.5$ strongly argues against scenarios in which observed LBGs cohabit their 
halos with a large number of similar but ``dormant'' systems whose starbursts 
have faded or not yet begun.  As a useful empirical result, we find that the 
observed line/continuum ratio for cB58 is similar to those of high--redshift 
systems with quite different dust luminosities and nuclear activity levels.  
Finally, we report the detection of a second source close to the position of 
the cD elliptical in the $z = 0.37$ lensing cluster, which may be nonthermal 
continuum emission from the cD or CO line emission from a hitherto unknown 
background galaxy at $z \sim 1.48$ or $z \sim 2.73$.
\end{abstract}


\section{Introduction} \label{s-intro}

The Lyman break technique \citep{stei92} has proved to be extremely efficient 
in identifying large numbers of actively star--forming galaxies at $z \geq 2$
\citep{stei99,stei03}.  While its selection criteria require only that 
Lyman break galaxies (LBGs) have rest--UV colors comparable to those of 
UV--bright starbursts at $z = 0$ (e.g., Meurer, Heckman, \& Calzetti 1999), 
the method has proved to identify high--redshift systems that resemble these 
local counterparts in other respects as well.  The two populations have 
similar bolometric surface brightnesses \citep{meur97}, UV--through--optical 
spectral energy distributions \citep{elli96,sawi98}, and blueshifted 
interstellar absorption from outflowing gas \citep{pett98,pett01}.  The 
primary difference is that LBGs appear to be forming stars at rates which are 
larger in proportion to their larger physical sizes \citep{meur97}.  Even the 
most careful studies of LBGs at relatively short wavelengths, however, provide 
only a partial picture of their important physical parameters.  Although 
population synthesis models (Sawicki \& Yee 1998; Papovich, Dickinson, \& 
Ferguson 2001; Shapley et al. 2001) can reconstruct their past star formation 
histories, and rest--UV fluxes can be used to derive their present star 
formation rates \citep{mada96,stei99} modulo uncertain corrections for 
extinction (Ouchi et al. 1999; Peacock et al. 2000; Adelberger \& Steidel 
2000; Chapman et al. 2000; Seibert, Heckman, \& Meurer 2002; Nandra et al. 
2002; Webb et al. 2003), ignorance of their molecular gas contents makes 
it impossible to predict LBGs' {\it future} courses of star formation.  
Moreover, estimates of LBGs' total dynamical masses remain uncertain insofar 
as they are based solely on observations of nebular emission lines: ionized 
gas is more likely than molecular gas to reflect non--gravitational influences 
\citep{heck98}, and less likely to trace the full rotation curve in local 
starbursts \citep{lehn96}.  The detection of molecular gas in one or more LBGs 
is therefore of great interest in complementing the extensive work in the rest 
UV and optical.

Within the LBG population, the brightest observable object is the 
gravitationally lensed system MS\,1512--cB58 (hereafter cB58: Yee et al. 
1996). Thanks to magnification of a background $z = 2.7$ source into a 
fold arc by the $z = 0.37$ cluster MS\,1512+36, this galaxy has permitted 
detailed studies in the rest--frame UV \citep{elli96,pett00,pett02} and 
optical \citep{bech97,tepl00}, as well as detections in the rest--frame near 
\citep{bech98} and far infrared (Baker et al. 2001; van~der~Werf et al. 2001; 
cf. Sawicki 2000).  Two initial searches for CO(3--2) emission from cB58  
were unsuccessful \citep{fray97,naka97}; notwithstanding cB58's sub--solar 
metallicity, these raised the question of how stars could be forming at such a 
seemingly high rate out of such a seemingly small reservoir of molecular gas.  
Our measurement of a surprisingly low $322\,{\rm \mu m}$ flux density 
($\lambda_{\rm obs} = 1200\,{\rm \mu m}$: Baker et al. 2001), implying a lower 
UV extinction and total star formation rate than those assumed by 
\citet{fray97}, suggested that molecular line emission might also be weaker 
than previously expected.  The measurement by \citet{tepl00} of nebular 
emission at a redshift apparently outside the bandpass of the \citet{naka97} 
observations strengthened our conviction that a new and more sensitive search 
for molecular gas in cB58 was warranted.

In this paper, we report our successful detection of CO(3--2) emission from 
cB58-- in fact, the first detection of molecular line emission from any 
LBG.  \S \ref{s-obs} describes our acquisition and reduction of the data. 
\S \ref{s-cb58} discusses our measurement of the CO(3--2) line parameters 
for cB58 and our derivations of molecular gas and dynamical masses.  \S 
\ref{s-x} presents our detection of an additional source ``X'' in the field, 
close to the position of the cD elliptical in the $z = 0.37$ lensing cluster.  
We consider the likeliest origin of this emission to be nonthermal 
continuum emission from the cD or CO line emission from a background galaxy, 
although neither explanation is entirely satisfactory.  \S \ref{s-disc} lays 
out a coherent picture of the past, present, and future of star formation in 
cB58, discusses the implications for our understanding of the $z \sim 3$ Lyman 
break population as a whole, and assesses source--to--source variation in the 
millimeter line/continuum ratio at high redshift.  \S \ref{s-conc} 
summarizes our conclusions.  Throughout this paper, we will assume $H_0 = 
70\,h_{0.7}\,{\rm km\,s^{-1}\,Mpc^{-1}}$ and a flat $\Omega_\Lambda = 0.7$ 
cosmology.\footnote{For cB58 at $z = 2.729$, this cosmology gives $D_{\rm L} = 
22.7\,h_{0.7}^{-1}\,{\rm Gpc}$ and $D_{\rm A} = 1.63\,h_{0.7}^{-1}\,{\rm 
Gpc}$.  For the lensing cluster at $z = 0.37$, $D_{\rm L} = 
1.98\,h_{0.7}^{-1}\,{\rm Gpc}$ and $D_{\rm A} = 1.06\,h_{0.7}^{-1}\,{\rm 
Gpc}$.}  The correction of observed fluxes for lensing magnification generally 
makes use of the most satisfactory model MVIa of Seitz et al. (1998 = S98).  
In this interpretation, all of the background rest--UV source 
is magnified by a factor 2.39 into a counterimage (A2), while a lesser 
fraction ($f_{\rm arc} \sim 0.54 - 0.67$) of the background source is 
magnified by a larger factor ($\propto f_{\rm arc}^{-1} \sim 59 - 48$) into 
the cB58 fold arc, giving an arc/intrinsic brightness ratio of ${\cal M}_{\rm 
UV} \simeq 31.8$.

\section{Observations}\label{s-obs}

We observed cB58 with the IRAM Plateau de Bure Interferometer (PdBI: 
Guilloteau et al. 1992) for 18 intervals from April 2001 through August 2002.  
The array included between four and six 15\,m diameter antennas; during our 
observations, these were arranged in the compact D configuration whose 15 
baselines range from 24 to 82\,m in length.  Each antenna is equipped with 
both 3\,mm and 1\,mm SIS receivers that can be used simultaneously; these were 
tuned in single--sideband mode at 3\,mm for 11 of the 18 observing epochs 
(encompassing $\sim 60\%$ of the data), and in double--sideband mode for the 
remaining 3\,mm and all 1\,mm data.  Receiver temperatures were 60--80\,K, 
and corrected outside the atmosphere yielded typical system temperatures of 
$\sim 120\,{\rm K}$ and $\sim 400\,{\rm K}$ at 3\,mm and 1\,mm, respectively.
For our observations of cB58, we adopted a phase center ($\alpha_{\rm J2000} = 
15^{h} 14^{m} 22.25^{s}$ and $\delta_{\rm J2000} = +36^{\circ} 36\arcmin 
24.40\arcsec$) $0.6\arcsec$ east of the coordinates listed in NED 
\citep{abra98}.  For the $z_{\rm H\,II} = 2.7290 \pm 0.0007$ reported by 
\citet{tepl00}, the CO(3--2) and CO(7--6) rotational transitions are 
redshifted to 92.73\,GHz and 216.32\,GHz.  We deployed four correlator 
modules-- giving a total of 580\,MHz of continuous bandwidth at $2.5\,{\rm 
MHz\,channel^{-1}}$ resolution-- across each of the two lines; both were 
observed in the lower sideband relative to the reference frequency.  After 
reducing the 2001 subset of the data, we shifted the 3\,mm receiver tuning to 
92.83\,GHz, in order to obtain more symmetric coverage of the already apparent 
CO(3--2) line at $z_{\rm CO} < z_{\rm H\,II}$ (see \S \ref{ss-cbpara}).  As a 
result, only the central $\sim 480\,{\rm MHz}$ of 3\,mm bandwidth containing 
data from all 18 observing epochs has been used to produce the line and 
continuum maps discussed below.

We calibrated the data using the CLIC routines in the IRAM GILDAS package 
\citep{guil00}.  Phase and amplitude variations within each track were 
calibrated out by interleaving observations of 3C345 (at a separation $\Delta 
\theta = 17.7^\circ$ from cB58 on the sky) every 30 minutes; we also observed 
the weaker but closer ($\Delta \theta = 1.9^\circ$) quasar 1504+377 to permit 
a check of our astrometric uncertainty (see \S \ref{ss-cbpara} below).  
Passband calibration used 3C345 or another bright quasar.  The overall flux 
scale for each epoch was set by comparing 3C345 with 3C273 and/or with model 
sources CRL618 and MWC349.  Reference fluxes for 3C273 were extracted from an 
archive of dedicated monitoring data from the PdBI and the IRAM 30\,m 
telescope.  We estimate the accuracy of our final flux scales to be $\sim 
15\%$.  

Before constructing $uv$ tables from our calibrated data, we smoothed them to 
a frequency resolution of 5\,MHz (i.e., velocity resolutions $16.2\,{\rm 
km\,s^{-1}}$ for the CO(3--2) line and $6.9\,{\rm km\,s^{-1}}$ for the 
CO(7--6) line).  After editing the data for quality in the DIFMAP package 
\citep{shep97}, we were left with the (on--source, six--telescope array) 
equivalents of 32 hours of 3\,mm data and 25 hours of 1\,mm data.  We inverted 
the data cubes with natural weighting (but no deconvolution) using the IMAGR 
task in the NRAO AIPS package \citep{vanm96}.  At 3\,mm, the lower--sideband 
data yield a synthesized beam of $8.2\arcsec \times 4.8\arcsec$ at P.~A. 
$73^\circ$, and a noise level per $16.2\,{\rm km\,s^{-1}}$ channel that 
increases with velocity (decreases with frequency) from 0.8 to $1.1\,{\rm 
mJy\,beam^{-1}}$ across the 480\,MHz bandpass.  The upper--sideband 3\,mm data 
from the epochs with double--sideband tuning give a synthesized beam of 
$8.2\arcsec \times 4.6\arcsec$ at P.~A. $75^\circ$, and a sensitivity to 
95.7\,GHz continuum emission of $0.24\,{\rm mJy\,beam^{-1}}$.  At 1\,mm, the 
lower--sideband data give a synthesized beam of $3.0\arcsec \times 2.1\arcsec$ 
at P.~A. $91^\circ$, and a noise level per $6.9\,{\rm km\,s^{-1}}$ channel of 
$5.3\,{\rm mJy\,beam^{-1}}$ across the entire 580\,MHz bandpass.  When both 
1\,mm sidebands are averaged together, the sensitivity to 217.8\,GHz 
continuum emission is $0.41\,{\rm mJy\,beam^{-1}}$ for the same synthesized 
beam.

\section{CO(3--2) emission from cB58} \label{s-cb58}

\subsection{Line parameters} \label{ss-cbpara}

\begin{figure}
\plotone{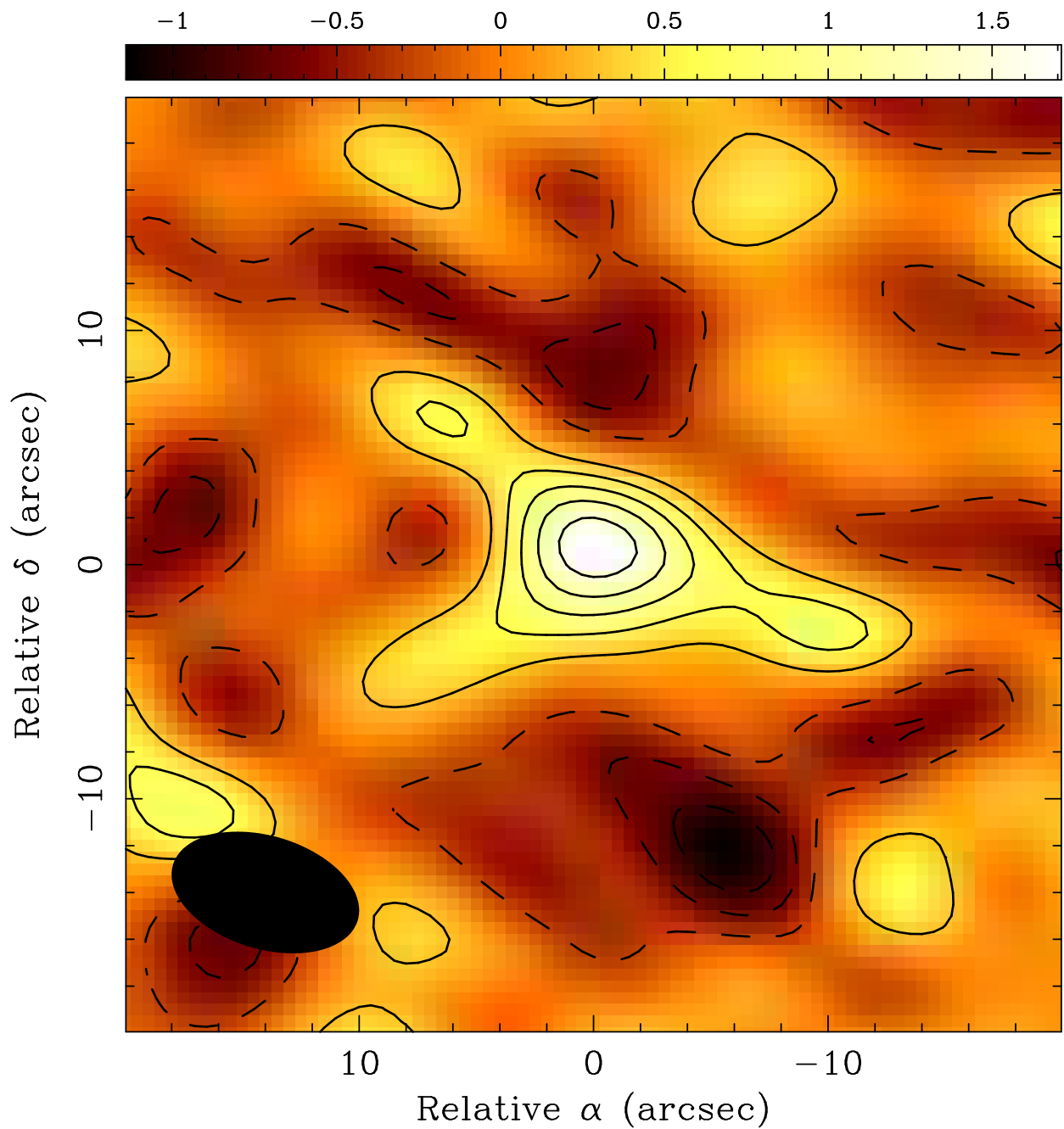}
\caption{Zeroth moment of the core of the CO(3--2) line in cB58.  
Coordinates are relative to the PdBI pointing center listed in Table 
\ref{t-pos}.  Emission has been averaged (without blanking) over a 60\,MHz 
($\sim 194.0\,{\rm km\,s^{-1}}$) window centered at 92.789\,GHz ($185.9\,{\rm 
km\,s^{-1}}$ blueward of $z_{\rm H\,II}$), i.e., over four of the $48.5\,{\rm 
km\,s^{-1}}$ channels in Figure \ref{f-cbspec} below.  Contours are multiples 
of $0.3\,{\rm mJy\,beam^{-1}}$ for a synthesized beam of $8.2\arcsec \times 
4.8\arcsec$ at P.~A. $73^\circ$ (shown at lower left); negative contours are 
dashed.  Interpreted as line emission, the $1.7 \pm 0.3\,{\rm mJy\,beam^{-1}}$ 
peak at the phase center corresponds to a CO(3--2) flux of $0.33 \pm 
0.06\,{\rm Jy\,km\,s^{-1}}$.  No primary beam correction has been applied.
\label{f-cbmap}}
\end{figure}

Cursory inspection of the 3\,mm data cube reveals the presence of a line 
emission source close to the phase center with velocity width $\sim 200\,{\rm 
km\,s^{-1}}$.  To quantify the significance and parameters of this source, we 
convolved the $16.2\,{\rm km\,s^{-1}}$ resolution cube with a twelve--channel 
boxcar filter.  The resulting cube contains $194.0\,{\rm km\,s^{-1}}$ 
channels, each of which overlaps its immediate neighbors by $177.8\,{\rm 
km\,s^{-1}}$.  The strongest source anywhere in this smoothed cube is 
shown in Figure \ref{f-cbmap}: neglecting primary beam correction, it is a 
$1.7 \pm 0.3\,{\rm mJy\,beam^{-1}}$ peak centered at 92.79\,GHz (i.e., $\sim 
190\,{\rm km\,s^{-1}}$ blueward of the nominal $z_{\rm H\,II}$).  A 
two--dimensional Gaussian fit in the image plane indicates that it is 
spatially unresolved; Table \ref{t-pos} lists the position derived from a fit 
that fixes the dimensions of the Gaussian to those of the synthesized beam.  
Two main lines of evidence argue that this $1.7 \pm 0.3\,{\rm mJy\,beam^{-1}}$ 
peak is in fact the $0.33 \pm 0.06\,{\rm Jy\,km\,s^{-1}}$ core of a CO(3--2) 
emission line from cB58:
\begin{enumerate}

\item The position of the source.  We can accurately register optical source 
positions into the radio reference frame for this field by identifying the 
($17\sigma$) 7\,mJy radio core in the $0.4\arcsec$ resolution 6.1\,cm map of 
\citet{laur97} with the central cD galaxy in the optical catalogue of 
\citet{abra98}.\footnote{NED positions for sources in this field have been 
calculated by identifying the cD with a less precisely located 21\,cm source 
(Becker, White, \& Helfand 1995).}  Under this assumption, the position of the 
optical fold arc agrees with the position of the putative CO(3--2) detection 
to within the latter's astrometric uncertainties (see Table \ref{t-pos}).  
Following \citet{down99}, we have calculated these uncertainties as the sum of 
three terms in quadrature:
\begin{enumerate}

\item The $\sim 0.01\arcsec$ absolute uncertainty of the PdBI reference frame; 
this stems from the fundamental uncertainties of quasar positions.

\item The relative uncertainty introduced by using a calibrator (3C345) that 
is offset by $\Delta \theta \simeq 17.7^\circ$ from cB58.  In principle, this 
uncertainty is about $0.01\arcsec$ per degree of separation, i.e., $\sim 
0.17\arcsec$.  However, because any error is expected to vary smoothly with 
position, we can exploit the fact that the quasar 1504+377 (only $1.9^\circ$ 
away from cB58, but $19.0^\circ$ away from 3C345) lies within $0.125\arcsec$ 
of the phase center in a map also calibrated by 3C345.  For cB58, lying 
between 3C345 and 1504+377, we can then impose the somewhat tighter limit of 
$(17.7/19.0) \times 0.125\arcsec \simeq 0.12\arcsec$.

\item The accuracy of our centroiding, which AIPS indicates is $\pm 0.6\arcsec$
in right ascension and $\pm 0.4\arcsec$ in declination.  This term 
dominates the other two.

\end{enumerate}

\item The redshift of the line.  \citet{pett00,pett02} have arrived at 
estimates of the systemic redshift of cB58 from rest--UV photospheric 
absorption features that are blueward of the \citet{tepl00} $z_{\rm H\,II}$ by 
$\sim 110-180\,{\rm km\,s^{-1}}$ in the rest frame of the source (see Table 
\ref{t-z}).  From fits to the CO line profile (see below), we estimate a 
$z_{\rm CO} = 2.7265^{+0.0004}_{-0.0005}$, whose $\sim 200\,{\rm km\,s^{-1}}$ 
blueshift relative to $z_{\rm H\,II}$ agrees fairly well with the rest--UV 
results.  We do not view the consistent difference from $z_{\rm H\,II}$ as 
necessarily astrophysically significant; if ${\rm H\alpha}$ were to 
arise largely in a wind, for example, we would expect $z_{\rm H\,II}$ to be 
{\it lower} than $z_{\rm CO}$, contrary to what is seen.  The discrepancy may 
rather reflect simple uncertainties in wavelength calibration.

\end{enumerate}
Three further consistency checks offer no evidence against this proposed 
identification.  First, our detection is consistent with the \citet{fray97} 
$3\sigma$ upper limit of $0.86\,{\rm Jy\,km\,s^{-1}}$ for a $194.0\,{\rm 
km\,s^{-1}}$ velocity width.  Second, we see no sign of a velocity gradient in 
the CO(3--2) data cube; this negative result is consistent with the absence of 
any shear along the arc in rest--UV observations of higher spatial resolution 
\citep{pett02}.  Finally, the lack of obvious 1\,mm continuum emission at the 
position of the putative CO(3--2) line (above a $3\sigma$ upper limit of 
1.2\,mJy) agrees with the results of \citet{bake01}, who report a 250\,GHz 
continuum flux density for cB58 of 1.06\,mJy.  Adopting their suggested 
$T_{\rm d} = 33\,{\rm K}$ dust temperature and $\epsilon_\nu = 2$ dust 
emissivity index, the 217.8\,GHz flux density expected in our PdBI map would 
be only 0.68\,mJy (see also \S \ref{ss-xor}).

\begin{figure}
\includegraphics{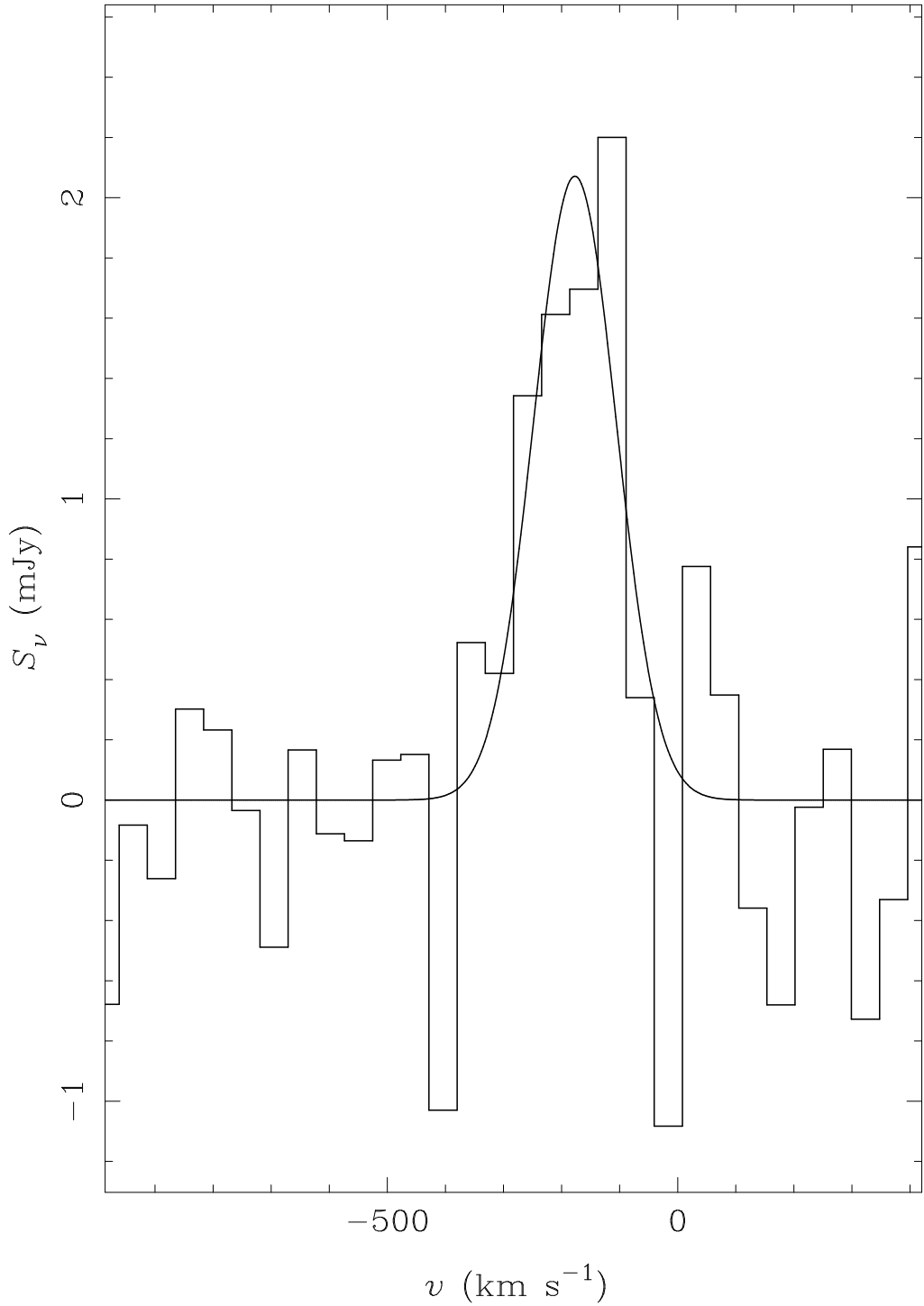}
\vspace{7.5in}
\caption{CO(3--2) spectrum of cB58 with a Gaussian fit overlaid.  Velocity 
channels are independent and $48.5\,{\rm km\,s^{-1}}$ wide; the zero point of 
the velocity axis corresponds to $z_{\rm H\,II} = 2.7290$ derived by 
\citet{tepl00}.  Gaussian $\sigma_{\rm CO} = {\rm FWHM_{CO}}/2.355 = 74 
\pm 18\,{\rm km\,s^{-1}}$.
\label{f-cbspec}}
\end{figure}

Given this identification, we proceed to calculate the parameters of the 
CO(3--2) emission in cB58.  While Figure \ref{f-cbmap} gives a fair estimate 
of the flux in the core of the line ($\sim 0.33\,{\rm Jy\,km\,s^{-1}}$), it 
fails to account for additional emission in the line wings extending blueward 
and redward of this particular $194.0\,{\rm km\,s^{-1}}$ chunk of the 
spectrum.  We have therefore fit Gaussians to the line profile through the 
brightest pixel in Figure \ref{f-cbmap}, as extracted from versions of the 
original $16.2\,{\rm km\,s^{-1}}$ cube that have been rebinned to $48.5\,{\rm 
km\,s^{-1}}$ resolution and corrected for primary beam attenuation (e.g., 
Figure \ref{f-cbspec}).  To estimate the uncertainties in the line widths and 
intensities, we have performed these fits on spectra smoothed from 16.2 to 
$48.5\,{\rm km\,s^{-1}}$ with all three possible choices of rebinning (i.e., 
averaging channels \{2--4, 5--7, ...\} and \{3--5, 6--8, ...\} as well as 
\{1--3, 4--6, ...\}).  In addition to $z_{\rm CO} = 2.7265$, we estimate a 
total line flux $F_{\rm CO(3-2)} = 0.37 \pm 0.08\,{\rm Jy\,km\,s^{-1}}$ and a 
velocity dispersion $\sigma_{\rm CO} = 74 \pm 18\,{\rm km\,s^{-1}}$ 
(equivalently, ${\rm FWHM}_{\rm CO} = 174 \pm 43\,{\rm km\,s^{-1}}$).

In the 1\,mm data cube, we have summed putative CO(7--6) channels over 
a $194.0\,{\rm km\,s^{-1}}$ velocity interval within $2.3\,{\rm km\,s^{-1}}$ 
of that integrated to produce Figure \ref{f-cbmap}.  At the position of the 
CO(3--2) peak, no CO(7--6) emission is seen; the nominal line flux is $F_{\rm 
CO(7-6)} = 0.15 \pm 0.21\,{\rm Jy\,km\,s^{-1}}$.  For a fair estimate of the 
CO(7--6)/CO(3--2) integrated intensity ratio ${\cal R}_{7/3}$, however, we 
must allow for the possibility that relative weakness in the CO(7--6) line 
can partly result from a relative lack of $uv$ sampling at short $r_{uv} = 
B/\lambda_{\rm obs}$, $B$ denoting the projected baseline length.  We 
have therefore truncated both of the original CO(3--2) and CO(7--6) $uv$ 
datasets below an identical minimum $r_{uv} = 1.1 \times 10^4$, and reinverted 
them with natural weighting.  After correcting for primary beam attenuation 
and convolving the new CO(7--6) map to the $7.2\arcsec \times 4.1\arcsec$ 
(P.~A. $80.9^\circ$) beam of the new CO(3--2) map, we take their quotient.  In 
flux units, the line ratio in a small region centered on the position of the 
CO(3--2) line detection is then $0.16 \pm 0.04$.  Translated to brightness 
temperature units, this yields a nominal integrated intensity ratio
\begin{equation}
{\cal R}_{7/3} \equiv {\frac {I_{\rm CO(7-6)}}{I_{\rm CO(3-2)}}} = 
\Big({\frac 3 7}\Big)^2\,\Big({\frac {F_{\rm CO(7-6)}}{F_{\rm CO(3-2)}}}\Big) 
= 0.030 \pm 0.007
\end{equation}
The lack of CO(7--6) emission in a noisy 1\,mm map means that this value 
may not be a particularly meaningful upper limit.  However, we can safely 
conclude that there is no evidence for large amounts of warm, dense molecular 
gas in cB58 like those observed in the submillimeter galaxy SMM\,J14011+0252 
at a similar redshift \citep{down03}.

To evaluate whether the width and intensity of the CO(3--2) line in cB58 are 
consistent with what we would expect for a relatively unobscured starburst, 
we must compare it to the parameters of CO(3--2) emission observed in 
appropriate $z = 0$ analogues.  \citet{maue99} report single--pointing 
CO(3--2) observations of a wide range of galaxy types; however, a better 
comparison sample is that of \citet{dumk01}, who have mapped the CO(3--2) 
line across the disks of twelve nearby starburst and star--forming galaxies.  
The most revealing quantity for comparison with cB58 is the beam--diluted 
brightness temperature $f_{\rm A} T_{\rm b}$, where $f_{\rm A}$ is the areal 
filling factor of a galaxy's molecular clouds and $T_{\rm b}$ is the 
rest--frame brightness temperature of a single cloud.  Following 
\citet{down03}, we can write the ``unlensed'' CO luminosity in terms of both 
observed and rest--frame quantities:
\begin{eqnarray}
{\frac {L^\prime_{\rm CO}}{\rm K\,km\,s^{-1}\,pc^2}} & = & 
362\,\Big({\frac {F_{\rm CO}}{\rm Jy\,km\,s^{-1}}}\Big)\,\Big({\frac 
{\lambda_{\rm obs}}{\rm mm}}\Big)^2\,\Big({\frac {D_{\rm A}}{\rm 
Mpc}}\Big)^2\,\Big({\frac {\cal M_{\rm CO}}{1}}\Big)^{-1}\,(1+z) 
\label{e-lco1} \\
 & = & 23.5\,\Big({\frac {T_{\rm b}}{\rm K}}\Big)\,\Big({\frac {\Delta v}{\rm 
km\,s^{-1}}}\Big)\,\Big({\frac {f_{\rm V}}{1}}\Big)\,\Big({\frac {f_{\rm 
A}}{1}}\Big)\,\Big({\frac {\Delta \Omega_{\rm CO}}{\rm arcsec^2}}\Big)\,
\Big({\frac {D_{\rm A}}{\rm Mpc}}\Big)^2
\label{e-lco2}
\end{eqnarray}
where $\Delta \Omega_{\rm CO}$ is the solid angle in the source plane, $f_{\rm 
V}$ is a velocity filling factor generally assumed to be unity due to the 
smoothness of extragalactic CO line profiles (e.g., Dickman, Snell, \& 
Schloerb 1986), and ${\cal M}_{\rm CO}$ is the factor by which the intrinsic 
CO(3--2) emission is magnified due to lensing.  Solving for $f_{\rm A} T_{\rm 
b}$, we find
\begin{equation}
{\frac {f_{\rm A}\,T_{\rm b}}{\rm K}} = 15.4\,\Big({\frac {F_{\rm CO}}{\rm 
Jy\,km\,s^{-1}}}\Big)\,\Big({\frac {\lambda_{\rm obs}}{\rm mm}}\Big)^2\,
\Big({\frac {\Delta v}{\rm km\,s^{-1}}}\Big)^{-1}\,\Big({\frac {\Delta 
\Omega_{\rm CO}}{\rm arcsec^2}}\Big)^{-1}\,\Big({\frac {\cal M_{\rm CO}}{1}}
\Big)^{-1}\,(1+z)
\label{e-ft}
\end{equation}
independent of cosmology.  For the \citet{dumk01} sample, we can set ${\cal 
M_{\rm CO}} = 1 + z = 1$ and $\lambda_{\rm obs} = 0.87\,{\rm mm}$; the 
CO(3--2) line flux, line width, and solid angle\footnote{Computed as $\pi D_1 
D_2/4$ from source diameters $D_1$ and $D_2$ in arcseconds.} of the dominant 
``central peak'' they determine for each system are listed in their Table 3.  
The resulting \{$f_{\rm A} T_{\rm b}$\} range from 0.4 to 5.4\,K.  For cB58, 
we can assume that the regions producing CO(3--2) line and rest--UV 
continuum emission (i.e., molecular clouds and the stars forming from them) 
are very similar, so that ${\cal M}_{\rm CO} \simeq {\cal M}_{\rm UV} \simeq 
31.8$.  By the same token, since model MVIa of S98 implies that the background 
LBG lensed into the cB58 arc has semi--major axes $a_{\rm UV} \times b_{\rm 
UV} \simeq 0.25\arcsec \times (0.25/2.39)\arcsec$, we can assume $\Delta 
\Omega_{\rm CO} \simeq \Delta \Omega_{\rm UV} = 0.082\,{\rm arcsec}^2$.  For 
$\lambda_{\rm obs} = 3.23\,{\rm mm}$ and the line flux and FWHM derived above, 
Equation \ref{e-ft} then implies cB58 has $f_{\rm A} T_{\rm b} \simeq 
0.5^{+0.3}_{-0.2}\,{\rm K}$.  This value falls near those seen in the centers 
of M\,51 (0.4), NGC\,278 (0.6), and NGC\,4631 (0.8) in the \citet{dumk01} 
sample.

The agreement of $f_{\rm A} T_{\rm b}$ estimates for cB58 and three nearby 
star--forming galaxies is strictly empirical-- in particular, we have not (yet)
invoked any relationship between $L^\prime_{\rm CO(3-2)}$ and molecular gas 
mass-- and has three general implications.  First, the consistency gives 
tentative support to our assumption that the system's rest--UV continuum and 
CO line emission are identically lensed; there is no order--of--magnitude 
discrepancy that can only be accounted for by differential lensing.  Second, 
it appears entirely plausible that cB58 contains molecular clouds whose 
distribution (i.e., $f_{\rm A}$) and typical internal conditions (i.e., 
$T_{\rm b}$) overlap with the ranges seen in local star--forming galaxy 
populations.  A stronger formulation of this point is not warranted, given 
that the relative weighting of $f_{\rm A}$ and $T_{\rm b}$ in the product of 
both could still differ at high and low redshift.  Finally, the molecular ISM 
in cB58 does not appear to resemble that of an ultra--luminous infrared galaxy 
(ULIRG) merger.  Five of the six ULIRG molecular disks from which 
\citet{lise96} detect CO(3--2) emission have CO(1--0) line widths measured by 
\citet{solo97} and intrinsic CO(1--0) and/or CO(2--1) source sizes inferred by 
\citet{down98}.  Assuming that these systems' $\Delta v$ and $\Delta 
\Omega_{\rm CO}$ are similar for all CO transitions, we estimate values for 
their $f_{\rm A} T_{\rm b}$ in the range $16-47\,{\rm K}$.  By the same token, 
the parameters of CO(3--2) emission in the lensed IR--luminous galaxy 
F10214+4724 ($z = 2.3$) mapped by Downes, Solomon, \& Radford (1995) imply 
$f_{\rm A} T_{\rm b} \simeq 10\,{\rm K}$-- a factor of 20 larger than in cB58. 

\subsection{Molecular gas mass} \label{ss-cbmgas}

The CO(1--0) line luminosity is well established as a proxy for the total gas 
mass gravitationally bound in a galaxy's molecular clouds (Dickman et al. 
1986; Solomon \& Barrett 1991; Wolfire, Hollenbach, \& Tielens 1993).  In 
terms of a conversion factor $X_{\rm CO} \equiv N_{\rm H_2}/I_{\rm 
CO(1-0)}\,{\rm cm^{-2}\,(K\,km\,s^{-1})^{-1}}$,
\begin{equation}
{\frac {M_{\rm gas}}{M_\odot}} = 4.4\,\Big({\frac {X_{\rm CO}}{2 \times 
10^{20}}}\Big)\,\Big({\frac {L^{\prime}_{\rm CO(1-0)}}{\rm K\,km\,s^{-1}\,
pc^2}}\Big)
\label{e-xfac}
\end{equation}
after including a factor of 1.36 to account for helium.  Although $X_{\rm CO} 
= (2-3) \times 10^{20}$ applies reasonably well for molecular clouds 
in the Galaxy and in the disks of nearby spirals \citep{stro88,solo91,scov91}, 
for galaxies with subsolar metallicity the universality of the conversion 
factor is a contested issue.  \citet{wils95} and Arimoto, Sofue, \& Tsujimoto 
(1996) argue that 
$X_{\rm CO}$ rises for lower--metallicity systems; averaging their recommended 
values for the metallicity of cB58, which we take to be $12 + {\rm log\,(O/H)} 
\simeq 8.39$ from \citet{tepl00}, would imply $X_{\rm CO} = 7.3 \times 
10^{20}$.  In contrast, \citet{roso03} find no systematic trend of $X_{\rm
CO}$ with metallicity in the range $0.25 \leq Z/Z_\odot \leq 1$ across the
disk of M\,33, while \citet{bola03} report $X_{\rm CO} = 7.0 \times 10^{20}$ 
for the $Z/Z_\odot = 0.1$ SMC, considerably smaller than the \citet{wils95} 
and \citet{arim96} relations would predict.  To reflect the range of these 
results, we adopt $X_{\rm CO} = (4.5 \pm 2.5) \times 10^{20}\,{\rm 
cm^{-2}\,(K\,km\,s^{-1})^{-1}}$ for cB58 in this paper.

Next, we must estimate cB58's integrated intensity ratio ${\cal R}_{3/1} 
\equiv I_{\rm CO(3-2)}/I_{\rm CO(1-0)}$.  \citet{dumk01} report values ranging 
from 0.4 to 1.4 in the centers, and from 0.5 to 1.0 in the disks, of their 
sample of twelve galaxies.  The centers of M\,51 and NGC\,278 are observed to 
have ${\cal R}_{3/1} = 0.5$ and 0.8, respectively; based on the analogy with 
cB58 in \S \ref{ss-cbpara} above, we adopt ${\cal R}_{3/1} = 0.65$ for cB58.  
Now combining Equations \ref{e-lco1} and \ref{e-xfac}, 
\begin{equation}
{\frac {M_{\rm gas}}{M_\odot}} = 5.5 \times 10^3\,\Big({\frac {X_{\rm CO}}{4.5 
\times 10^{20}}}\Big)\,\Big({\frac {{\cal R}_{3/1}}{0.65}}\Big)^{-1}\,
\Big({\frac {F_{\rm CO(3-2)}}{\rm Jy\,km\,s^{-1}}} \Big)\,\Big({\frac 
{\lambda_{\rm obs}}{\rm mm}}\Big)^2\,\Big({\frac {D_{\rm A}}{\rm Mpc}}\Big)^2 
\Big({\frac {\cal M_{\rm CO}} 1}\Big)^{-1}\,(1+z) \label{e-mgas}
\end{equation}
Again assuming ${\cal M}_{\rm CO} \simeq {\cal M}_{\rm UV}$, and including the 
uncertainty in the appropriate value of $X_{\rm CO}$, we estimate for cB58 
$M_{\rm gas} = 6.6^{+5.8}_{-4.3} \times 10^9\,h_{0.7}^{-2}\,M_\odot$.  We 
make no correction for the atomic mass component of cB58's ISM, although we 
note that in low--metallicity systems, a large fraction of the gas that is 
gravitationally bound into the ``molecular'' clouds traced by the CO lines 
(and thus readily available for star formation) may already be atomic in 
phase.  

cB58's molecular gas mass is comparable to those seen in the more gas--rich 
members of the local spiral population (e.g., Helfer et al. 2003), somewhat 
smaller than those seen in nearby ULIRGs (Bryant \& Scoville 1999; cf. Downes 
\& Solomon 1998), and up to an order of magnitude less than those inferred for 
submillimeter--selected systems at comparably high $z \geq 2.5$ 
\citep{fray98,fray99,neri03}.  If its molecular gas is spread over a disk of 
radius $r_{\rm CO} = 0.25\arcsec \sim 2.0\,h_{0.7}^{-1}\,{\rm kpc}$ in the 
source plane (see \S \ref{ss-cbmdyn} below), the corresponding mass surface 
density will be $\Sigma_{\rm gas} = 540^{+470}_{-350}\,M_\odot\,{\rm 
pc}^{-2}$.  Such surface densities are more reminiscent of the centers of 
gas--rich spirals \citep{helf03} than of ULIRGs, consistent with the 
similarity noted above between cB58 and the \citet{dumk01} sample in terms 
of $f_{\rm A} T_{\rm b}$.  Since the use of a single conversion factor $X_{\rm 
CO}$ in different galaxies relies on the assumption that their molecular cloud 
properties are reasonably similar \citep{dick86}, this consistency serves to 
validate our use of a (metallicity--corrected) Galactic $X_{\rm CO}$ to 
estimate $M_{\rm gas}$ in the first place.\footnote{Indeed, our use of a 
conversion factor exceeding the Galactic $X_{\rm CO}$ by a factor of 2.25 
implies that in terms of its mean molecular hydrogen column (${\bar N}_{\rm 
H_2} \propto X_{\rm CO} f_{\rm A} T_{\rm b}$), cB58 will more closely resemble 
the galaxies in the \citet{dumk01} sample whose beam--diluted brightness 
temperatures are more typical (i.e., not among the lowest observed).}

\subsection{Dynamical mass} \label{ss-cbmdyn}

Our spectrally resolved CO(3--2) detection means that we can estimate the 
dynamical mass of cB58 from its velocity width according to
\begin{equation}
{\frac {M_{\rm dyn}}{M_\odot}} = 1.1 \times 10^3\,\Big(\alpha\Big)\,
\Big({\frac {r_{\rm CO}}{\rm arcsec}}\Big)\,\Big({\frac{\sigma_{\rm CO}}
{\rm km\,s^{-1}}}\Big)^2\,\Big({\frac {D_{\rm A}}{\rm Mpc}}\Big)
\label{e-mdyn}
\end{equation}
Here $\alpha$ is a constant of proportionality whose value depends on the 
intrinsic geometry of cB58's molecular gas, and on how that geometry couples 
with the effects of lensing.  Since molecular gas is extremely dissipative 
and unlikely to remain molecular in phase if severely disturbed, it is more 
likely than not that the CO emission from cB58 arises in a rotating disk.  
Indeed, the S98 source plane reconstruction suggests that the 
fraction ($f_{\rm arc} \sim 0.6$) of the background LBG being magnified into 
the fold arc corresponds to half of such a disk, slightly inclined and with 
half--light radius $r_{\rm UV} = 0.25\arcsec$.  In this scenario, assuming 
that $r_{\rm CO} \simeq r_{\rm UV}$ and that the magnified half of the disk 
traces only half of the galaxy's full line width, we would infer $\alpha 
\simeq 4\,{\rm csc}^2 i$.  This corresponds to a dynamical mass $M_{\rm dyn} = 
1.0^{+0.6}_{-0.4} \times 10^{10}\,{\rm csc}^2 i\,h_{0.7}^{-1}\,M_\odot$ within 
a radius of $2.0\,h_{0.7}^{-1}\,{\rm kpc}$, with the uncertainties in $M_{\rm 
dyn}$ reflecting only the uncertainties in $\sigma_{\rm CO}$.

If we wish to avoid relying too heavily on the details of the source 
morphology reconstructed by S98, we can also consider the 
possibility that cB58's molecular gas has a spherical rather than disklike 
distribution.  In this scenario, $\alpha = 5$ (for a sphere of uniform 
density) and correction of the observed half--light radius of A2 for lensing 
yields a characteristic $r_{\rm CO} \simeq (0.25/\sqrt{2.39})\arcsec = 
0.16\arcsec$.  The implied dynamical mass is now a somewhat smaller $M_{\rm 
dyn} = 0.8^{+0.5}_{-0.4} \times 10^{10}\,h_{0.7}^{-1}\,M_\odot$ within a 
radius of $1.3\,h_{0.7}^{-1}\,{\rm kpc}$.  We view this estimate as less 
plausible than the result for a disklike geometry, however, for the {\it a 
priori} reasons noted above.

For either assumed gas configuration, we can compare $M_{\rm dyn}$ with the 
$M_{\rm gas}$ derived in \S \ref{ss-cbmgas} above to estimate the molecular
gas mass fraction $f_{\rm gas}$ in the center of cB58.  For a disk geometry, 
$f_{\rm gas} = 0.66^{+0.34}_{-0.52}\,{\rm sin}^2 i\,h_{0.7}^{-1}$ within a 
$2.0\,h_{0.7}^{-1}\,{\rm kpc}$ radius; for a spherical geometry, $f_{\rm gas} 
= 0.82^{+0.18}_{-0.64}\,h_{0.7}^{-1}$ within $1.3\,h_{0.7}^{-1}\,{\rm kpc}$.
Both scenarios imply a rather high gas mass fraction, although 
the uncertainties in both estimates (including ${\rm sin}^2 i$ for the disk 
case) make it impossible to exclude the possibility that $f_{\rm gas} \ll 
0.5$.  

\section{Emission from a second source} \label{s-x}

\subsection{X and its possible origins} \label{ss-xor}

\begin{figure}
\plotone{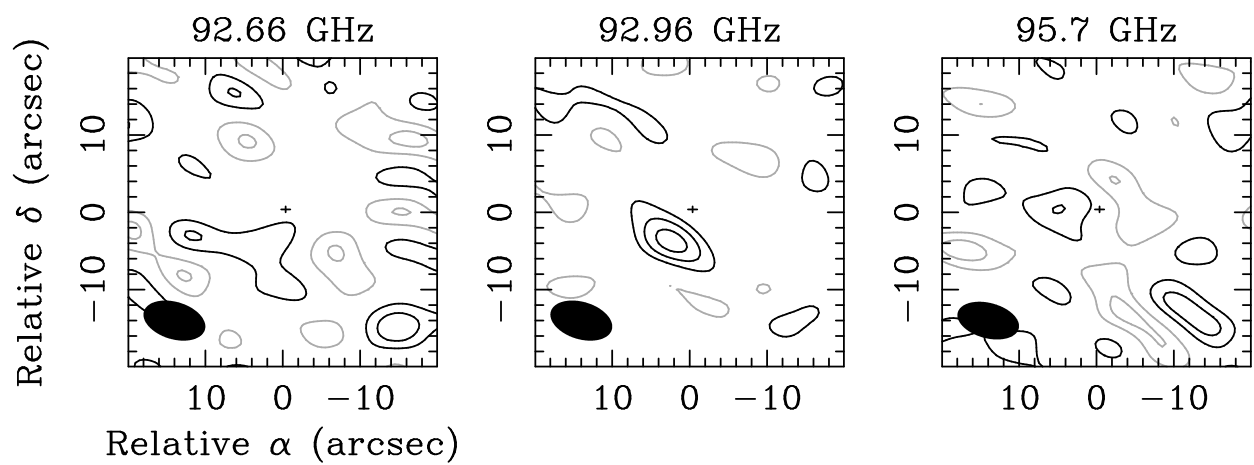}
\caption{Pseudo--continuum images at three different 3\,mm frequencies.  
Synthesized beams are plotted at lower left; crosses show $\pm 1\sigma$ 
positional uncertainties for the centroid of CO(3--2) emission from cB58.
Contours are multiples of $0.24\,{\rm mJy\,beam^{-1}}$, which corresponds to 
$1\sigma$ for the left and right panels and $1.5\sigma$ for the center panel; 
grey contours are negative.  The $5.4\sigma$ source seen to the southeast 
of cB58 in the center panel is called X in the text.  No primary beam 
correction has been applied.
\label{f-xmap}}
\end{figure}

Further examination of our 3\,mm data cube-- specifically, the velocity 
channels outside the range spanned by cB58's CO(3--2) emission-- reveals a 
second source in the same field.  Averaging over a 200\,MHz bandwidth centered 
at 92.96\,GHz (i.e., over a $645\,{\rm km\,s^{-1}}$ velocity range lying 
$523\,{\rm km\,s^{-1}}$ blueward of $z_{\rm CO}$ for cB58) yields the map 
shown in the center panel of Figure \ref{f-xmap}.  The new source (hereafter 
``X'') has an apparent flux density $0.87 \pm 0.16\,{\rm mJy}$ ($\pm 
0.13\,{\rm mJy}$ for the uncertain flux scale) after primary beam correction. 
A Gaussian fit gives the position listed in Table \ref{t-pos}, which coincides 
with the radio continuum core of the cD in the lensing cluster 
\citep{beck95,laur97} and is clearly distinct from 
the position of cB58's CO(3--2) centroid.\footnote{The tight astrometric 
constraints in a common radio reference frame also imply that cB58 cannot be 
the source of both the CO(3--2) emission discussed in \S \ref{s-cb58} and the 
6.1\,cm continuum emission mapped by \citet{laur97}.}  Puzzlingly, however, 
we do not detect comparably significant emission from X in a 140\,MHz 
bandwidth centered at 92.66\,GHz (nominally $0.46 \pm 0.24\,(\pm 0.07)\,{\rm 
mJy}$), or in our 95.7\,GHz continuum map (nominally $0.26 \pm 0.24\,(\pm 
0.04)\,{\rm mJy}$), as shown in the left and right panels of Figure 
\ref{f-xmap}.  X also fails to appear in our 217.8\,GHz continuum map 
(nominally $0.79 \pm 0.45\,(\pm 0.12)\,{\rm mJy}$).  

Since the noise is 50\% higher in our maps at 92.66\,GHz (due to a smaller 
bandwidth and the noise gradient described in \S \ref{s-obs}) and at 95.7\,GHz 
(due to the dearth of upper--sideband data) than at 92.96\,GHz, it remains 
possible that X is indeed a weak continuum source.  In this case, its 
positional coincidence with the cD suggests a likely nonthermal origin.  The 
cD's radio core has a 1.4\,GHz flux density $\sim 1\,{\rm mJy}$ at $5\arcsec$ 
resolution \citep{beck95}, and a 4.9\,GHz flux density $\sim 7\,{\rm mJy}$ at 
$0.4\arcsec$ resolution \citep{laur97}, meaning it either varies or has an 
unusually inverted slope ($\alpha^{4.9}_{1.4} \simeq -1.56$ for $S_\nu \propto 
\nu^{-\alpha}$).  In the \citet{dall00} sample of High Frequency Peakers 
(HFPs), $\alpha^{4.9}_{1.4}$ reaches values as extreme as $-1.72$, so the 
latter possibility is plausible.  Assuming the cD radio core is young, with 
$\alpha \sim -1.56$ below a turnover frequency $\nu_{\rm peak}$ as high as 
those of the most extreme HFPs ($\sim 22\,{\rm GHz}$) and $\alpha \sim +0.8$ 
above $\nu_{\rm peak}$ (e.g., Condon 1992), we would predict a 92.96\,GHz flux 
density $\sim 23\,{\rm mJy}$.  A less aggressively tuned SED could still 
account for all of the observed 0.87\,mJy from X.  This scenario would explain 
why no emission from X appears in our 217.8\,GHz continuum map, since a 
spectral index $\alpha = +0.8$ would imply $S_{217.8} \simeq 0.44\,{\rm mJy}$, 
i.e., at the level of the noise.

A less plausible continuum scenario accounts for X as thermal dust emission 
associated with star formation in the cD.  Analogous systems do exist; of the 
two $850\,{\rm \mu m}$ detections of $z \sim 0.25$ cDs by \citet{edge99}, for 
example, at least one (in Abell 1835) has an unambigous dust origin.  However, 
the tight constraints on emission from X at $\nu \gg 92.96\,{\rm GHz}$ make it 
unlikely that the cD in MS\,1512+36 is similar.  For the median dust 
temperature $T_{\rm d} = 58\,{\rm K}$ and (relatively flat) emissivity index 
$\epsilon_\nu = 1.35$ in the local starburst sample of \citet{yun02}, a $z = 
0.37$ galaxy with flux density 0.87\,mJy at 92.96\,GHz should have flux 
densities $\sim 14$, 22, and 65\,mJy at 1.4\,mm (217.8\,GHz), 1.2\,mm, and 
$850\,{\rm \mu m}$, respectively.  It is marginally possible that our PdBI 
map could have resolved out 14\,mJy of 217.8\,GHz emission from X.  It is 
completely inconceivable, however, that bolometer array observations of cB58 
could have missed such strong emission from a second source so close on the 
sky.  For the $11\arcsec$ and $15\arcsec$ (HPBW) beams of MAMBO at the IRAM 
30\,m telescope and SCUBA at the JCMT, a $4.8\arcsec$ separation translates to 
attenuation by factors of 0.59 and 0.75, respectively.  Accounting for this 
correction, \citet{bake01} should have measured $\sim 13\,{\rm mJy}$ at 
1.2\,mm for cB58, rather than the $1.06 \pm 0.35\,{\rm mJy}$ observed, while 
\citet{vand01} and \citet{sawi01} should have measured $\sim 49\,{\rm mJy}$ at 
$850\,{\rm \mu m}$, rather than their $4.2 \pm 0.9\,{\rm mJy}$ detection and 
3.9\,mJy $3\sigma$ upper limit.  Similar inconsistencies are encountered for 
any reasonable choices of $T_{\rm d}$ and $\epsilon_\nu$.

\begin{figure}
\includegraphics{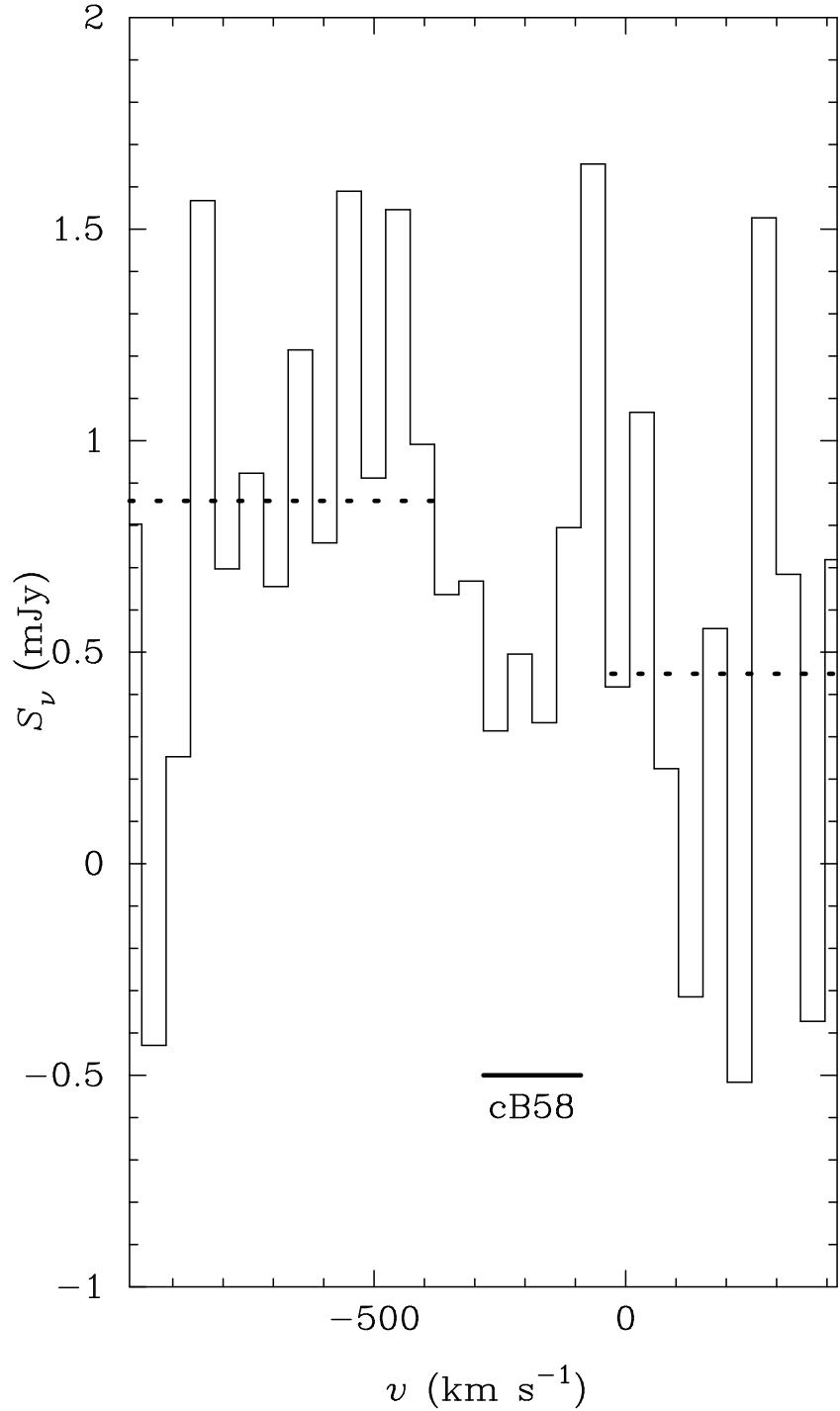}
\vspace{7.5in}
\caption{3\,mm ``spectrum'' of X with the same velocity channels as Figure 
\ref{f-cbspec}.  Dashed lines show the approximate windows averaged to 
produce the center and left panels of Figure \ref{f-xmap}.  Solid line at 
bottom shows the velocity range spanned by the core of cB58's CO(3--2) line 
as used to plot Figure \ref{f-cbmap}.
\label{f-xspec}}
\end{figure}

If we instead take Figure \ref{f-xmap} at face value and interpret X as 
an emission--line source, its positional coincidence with the cD must be 
essentially accidental.  Its irregular velocity ``profile'' (Figure 
\ref{f-xspec}) does not permit a detailed fit for an exact centroid or width, 
although it is clear that the latter must be comparable to the bandwidth 
smoothed to obtain the center panel of Figure \ref{f-xmap}.  Such a $\sim 
645\,{\rm km\,s^{-1}}$ velocity width is incompatible with the $246\,{\rm 
km\,s^{-1}}$ velocity dispersion for the cD required by the S98 lensing model; 
moreover, a rest frequency $\sim 127.35\,{\rm GHz}$ does not correspond to any 
strong recombination line (Towle, Feldman, \& Watson 1996) or molecular 
rotational transition.  We therefore adopt the reverse strategy of assuming X 
is one of the strongest extragalactic emission lines-- i.e., a CO rotational 
transition-- and asking whether an object at the corresponding redshift can 
provide a self--consistent explanation.  As CO(1--0) emission, X would 
necessarily come from a $z \simeq 0.24$ galaxy lying in front of the cluster.  
The only possible candidate is the foreground spiral ``S'' seen $\sim 
4.4\arcsec$ southeast of X, which S98 suggest has $0.25 \leq z_S \leq 0.37$ on 
the basis of its angular size and $V - R$ color.  However, its apparently 
face--on orientation makes a $\sim 645\,{\rm km\,s^{-1}}$ linewidth highly 
implausible, particularly at a radius $\sim 16.6\,h_{0.7}^{-1}\,{\rm kpc}$ 
from the center of a spiral that would otherwise appear to be CO--poor.

As CO(2--1) emission from a $z_X \sim 1.48$ galaxy or CO(3--2) emission from a 
(further) $z_X \sim 2.73$ galaxy, X would lie behind the lensing cluster and 
therefore undergo some degree of magnification.  A significant challenge 
for this scenario is that any background source producing an image within an 
arcsecond of the central cD should also produce a number of counterimages 
(over the same velocity range) elsewhere in the cluster field.  Velocity 
considerations alone mean X cannot be CO(3--2) emission from a fifth image in 
the W set of S98, since here the background source redshift lies only
$150\,{\rm km\,s^{-1}}$ from that of cB58 (Teplitz, Malkan \& McLean 2002).  
To explore the 
possibility that X is CO(2--1) or CO(3--2) emission from a hitherto unknown 
background source, we have implemented the S98 lensing model within the {\it 
gravlens} package \citep{keet01}.  As discussed in Appendix \ref{a-lens}, we 
have then assumed that X is one image of a lensed background point source and 
calculated where the corresponding counterimages should lie for both $z_X = 
1.48$ and $z_X = 2.73$.  The only configurations that do not predict a strong 
counterimage where none is observed require that X correspond to two merging 
images of opposite parity on opposite sides of a critical curve.  Because the 
details of such models depend sensitively on the exact shape of the lens mass 
profile, we view our particular results as largely speculative.  We do 
consider $z_X \simeq 2.73$ a more appealing possibility than $z_X \simeq 
1.48$, since we already know that cB58 and the background source producing the 
W image set lie at about the same redshift \citep{tepl02}.  The projected 
separation between the background galaxies lensed to produce cB58 and X would 
be only $3.1\arcsec \sim 25\,{\rm kpc}$ in this case; the presence of an 
additional, more obscured and gas--rich companion that could have helped 
trigger cB58's star formation recalls the multiple--component morphology of 
the LBG Westphal--MMD11 \citep{chap02}.  However, for either $z_X$, a lensed 
background source would need to have sufficiently strong dust extinction to 
avoid easy detection behind the cD in {\it HST} optical imaging, yet 
sufficiently weak dust emission to avoid easy detection in $1.2\,{\rm mm}$ and 
$850\,{\rm \mu m}$ bolometry.  This combination of properties requires a 
significant, perhaps fatal degree of fine tuning.

\subsection {Implications of X for parameters of cB58} \label{ss-ximp}

There are two ways in which the presence of source X within $4.8\arcsec$ of 
cB58 on the sky might lead us to adjust the properties of the emission we 
attribute to the LBG.  The first is the prospect that continuum (or broad 
line) emission from X has contaminated our measurements of the CO(3--2) 
emission line within the PdBI data cube.  This turns out not to be a problem.  
If we treat X as line emission, then the best estimate of its velocity 
centroid and width (see Figure \ref{f-xspec}) place it mostly blueward of 
cB58's CO(3--2) emission, and the positional offset suppresses any remaining 
contamination.  If we treat X as nonthermal continuum from the cD and 
subtract a model of the line--free channels from the $uv$ data before 
inverting a now ``line--only'' cube, we measure CO(3--2) line parameters for 
cB58 that are virtually identical to those derived in \S \ref{s-cb58} without 
the benefit of ``continuum subtraction'' (e.g., the peak in Figure 
\ref{f-cbmap} nominally increases from $1.7 \pm 0.3\,{\rm mJy}$ to $1.8 
\pm 0.3\,{\rm mJy}$).

The second possible problem is that if X is continuum emission from the 
cD or line emission from a dusty background source, there may be some 
contamination of the 1.2\,mm and $850\,{\rm \mu m}$ dust emission attributed 
to cB58 by \citet{bake01} and \citet{vand01}, respectively.  The amount of 
dust that could be associated with a lensed line source is difficult to 
quantify.  The possible contamination by nonthermal continuum from the cD is 
more tractable: assuming $\alpha = +0.8$ and accounting for the different 
MAMBO and SCUBA beam sizes, we would estimate that roughly 0.22\,mJy of each 
detection could have come from the cD.  This would represent $\sim 5\%$ of the 
$850\,{\rm \mu m}$ detection, but $\sim 22\%$ of the 1.2\,mm flux density.  To
the extent that this has occurred, the weaker dust emission attributed to cB58 
would support the conclusion of \citet{sawi01} that the galaxy's 
long--wavelength SED differs from those of local starbursts, although this 
difference would still be less dramatic than past (over)estimates of its 
rest--UV reddening have suggested \citep{bake01}.

\section{Discussion}\label{s-disc}

\subsection{Star formation in cB58}\label{ss-cbsf}

Previous studies of cB58's rest--UV spectroscopic properties have placed 
strong constraints on its most likely star formation history.  The 
underabundance of nitrogen and iron--peak elements inferred from a Keck/ESI 
spectrum (Pettini et al. 2002; cf. Teplitz et al. 2000) requires that star 
formation must have begun at some $t_{\rm SF(past)} \leq 300\,{\rm Myr}$, 
consistent with the lack of evidence for any older stellar population in  
rest--frame near--IR emission \citep{bech98}.  The initial population 
synthesis analysis by \citet{pett00} of their Keck/LRIS spectrum has been 
revisited by De~Mello, Leitherer, \& Heckman (2000) and \citet{leit01} using 
successive improvements 
of the original {\it Starburst99} package \citep{leit99}.  Overall, these 
three exercises find satisfactory agreement with the observed spectrum if cB58 
has been forming stars at a constant rate with a \citet{salp55} Initial Mass 
Function (IMF) up to $100\,M_\odot$ for the last $t_{\rm SF(past)} = 
20-100\,{\rm Myr}$.  We can use this picture of cB58's recent past to estimate 
its star formation rate in the present.  The $Z = 0.4\,Z_\odot$ {\it 
Starburst99} models for the above range in $t_{\rm SF(past)}$ predict
\begin{eqnarray}
{\frac {\rm SFR}{M_\odot\,{\rm yr^{-1}}}} & = & {\frac {L_{\rm H\alpha}}{1.34 
\times 10^{41}\,{\rm erg\,s^{-1}}}} \nonumber \\
 & = & 8.9 \times 10^{8}\,\Big({\frac {F_{\rm H\alpha}}{\rm 
erg\,cm^{-2}\,s^{-1}}}\Big)\,\Big({\frac {D_{\rm L}}{\rm Mpc}}\Big)^2\,
\Big({\frac {{\cal M}_{\rm H\alpha}}{1}}\Big)^{-1} \label{e-sfha}
\end{eqnarray}
where we have extended the Salpeter IMF to 
$0.1\,M_\odot$.  For cB58, \citet{tepl00} measure a Balmer decrement of 3.23 
and an ${\rm H\alpha}$ flux of $(12.56 \pm 0.37) \times 10^{-16}\,{\rm 
erg\,cm^{-2}\,s^{-1}}$; correcting the latter by a factor of 1.32 for LMC 
extinction \citep{howa83} and assuming ${\cal M}_{\rm H\alpha} \simeq 31.8$ 
implies ${\rm SFR} = (23.9 \pm 0.7)\,h_{0.7}^{-2}\,M_\odot\,{\rm yr^{-1}}$.  
Given that the distributions of UV and ${\rm H\alpha}$ emission in local 
starbursts are broadly similar \citep{cons00}, ${\cal M}_{\rm H\alpha} \simeq 
{\cal M}_{\rm UV}$ should indeed 
hold if the fraction of the background source being 
lensed is truly as high as $f_{\rm arc} \sim 0.6$.  The rate of star formation 
per unit area in cB58, $\Sigma_{\rm SFR} = 1.9 \pm 0.1\,M_\odot\,{\rm yr^{-1}\,
kpc^{-2}}$, appears within the uncertainties to be following a \citet{schm59} 
law dependence on $\Sigma_{\rm gas}$.  From the best--fit relation determined 
by \citet{kenn98},
\begin{equation} \label{e-sch}
{\frac {\Sigma_{\rm SFR}}{M_\odot\,{\rm yr^{-1}\,kpc^{-2}}}} = 2.5 \times 
10^{-4}\,\Big({\frac {\Sigma_{\rm gas}}{M_\odot\,{\rm pc^{-2}}}}\Big)^{1.4}
\end{equation}
we would have predicted cB58 to have $\Sigma_{\rm SFR} = 1.6^{+2.3}_{-1.2}\,
M_\odot\,{\rm yr^{-1}\,kpc^{-2}}$.

After continuous star formation at the current rate for the last 20--100\,Myr, 
cB58 should already have produced $M_{\rm stars} = (0.5 - 2.4) \times 
10^9\,h_{0.7}^{-2}\,M_\odot$ in the present episode.  For any such stellar 
mass, the implied baryonic mass-- $M_{\rm bary} \equiv M_{\rm gas} + M_{\rm 
stars} \simeq (7.1 - 8.9) \times 10^9\,M_\odot$-- is still safely less than 
the best estimate of the dynamical mass 
($M_{\rm dyn} = 1.0 \times 10^{10}\,{\rm 
csc}^2 i\,h_{0.7}^{-1}\,M_\odot$), an encouraging consistency check.  For 
several reasons, we consider a stellar mass ($\propto t_{\rm SF(past)}$) at 
the upper end of the allowed range to be the most plausible scenario.  First, 
estimates of small $t_{\rm SF(past)}$ (and low $M_{\rm stars}$) that involve 
rest--UV continuum colors (e.g., Ellingson et al. 1996; Pettini et al. 2000) 
entail higher levels of extinction than are indicated by several other lines 
of evidence \citep{bake01}.  Second, the difference $M_{\rm dyn} - M_{\rm 
bary}$ should be small, since (a) there is no evidence for any {\it very} old 
stellar population (see above), and (b) we do not expect large dark matter 
contributions to the total mass budget at $R \leq 2.0\,h_{0.7}^{-1}\,{\rm 
kpc}$.  Indeed, assuming $i = 90^\circ$ and $t_{\rm SF(past)} \simeq 140\,{\rm 
Myr}$ (all but indistinguishable from $t_{\rm SF(past)} = 100\,{\rm Myr}$ in a 
population synthesis model with continuous star formation) would leave $M_{\rm 
dyn} = M_{\rm bary}$ exactly.  We will therefore assume $t_{\rm SF(past)} 
\simeq 140\,{\rm Myr}$ and $M_{\rm stars} \simeq 3.4 \times 10^9\,M_\odot$ 
in what follows.  Third, if cB58 is a typical LBG in respects 
other than its fortuitous location behind a lensing cluster, we would expect 
its $M_{\rm stars}$ to be comparable to those inferred from population 
synthesis analyses of LBGs with similar intrinsic brightnesses.  Correcting 
the observed $K\arcmin = 17.83$ \citep{elli96} for lensing implies cB58 has an 
intrinsic $K\arcmin \simeq 21.59$.  32 of the 74 LBGs for which 
\citet{shap01} succeed in deriving SED fits have $K_{\rm s}$ within 0.5\,mag 
of this value.  These have a mean stellar mass of $4.0 \times 10^{10}\,
h_{0.7}^{-2}\,M_\odot$; a healthy five of the 32 have $M_{\rm stars} \leq 3.4 
\times 10^9\,h_{0.7}^{-2}\,M_\odot$, but only one has $M_{\rm stars} \leq 0.5 
\times 10^9\,h_{0.7}^{-2}\,M_\odot$.  Finally, following the argument 
of \citet{pett02}, $M_{\rm stars} \sim 3.4 \times 10^9\,M_\odot$ would also 
imply that cB58 has already turned a fraction $M_{\rm stars}/M_{\rm bary} \sim 
1/3$ of its baryons into stars, roughly the level required to account for the 
enrichment of cB58's ISM to the observed $Z = 0.4\,Z_\odot$.  In fact, the 
stellar/baryonic mass fraction is close enough to this threshold to reinforce 
the argument against previous episodes of star formation; these would tend to 
have elevated the galaxy's oxygen abundance above its observed value.

Besides charting its past history, we can predict how far into 
the future cB58 can continue forming stars at the present rate before it
exhausts its readily available fuel.  Combining Equations \ref{e-mgas} and 
\ref{e-sfha}, we have (independent of cosmology) a gas consumption timescale
\begin{eqnarray}
{\frac {t_{\rm gas}}{\rm Myr}} & = & 6.2 \times 10^{-12}\,\Big({\frac 
{F_{\rm CO(3-2)}}{\rm Jy\,km\,s^{-1}}}\Big)\,\Big({\frac {F_{\rm 
H\alpha}}{\rm erg\,cm^{-2}\,s^{-1}}}\Big)^{-1} \nonumber \\
 & & \times \Big({\frac {X_{\rm CO}}{4.5 \times 10^{20}}}\Big)\,\Big({\frac 
{{\cal R}_{3/1}}{0.65}}\Big)^{-1}\,\Big({\frac {\lambda_{\rm obs}}{\rm 
mm}}\Big)^2\,\Big({\frac {{\cal M}_{\rm H\alpha}}{{\cal M}_{\rm CO}}}\Big)\,(1 
+ z)^{-3}
\end{eqnarray}
For cB58, we find $t_{\rm gas} \simeq 280^{+280}_{-180}\,{\rm Myr}$, an 
eminently reasonable value.  In nearby spiral galaxies, using far--IR 
luminosity as a proxy for star formation rate, \citet{deve91} have showed that 
the median $t_{\rm gas}$ is $\sim 1\,{\rm Gyr}$.  In contrast, the most 
luminous local ULIRGs have more efficient star formation and $t_{\rm gas}$ 
approaching $\sim 100\,{\rm Myr}$ \citep{brya99}.  cB58, as befits a starburst 
of intermediate intensity, falls between these two extremes.

An important caveat in forecasting the future of star formation in cB58 
is that the remaining lifetime of the current episode, $t_{\rm SF(future)}$, 
may be less than $t_{\rm gas}$ if a large fraction of the galaxy's molecular 
gas reservoir is expelled by an energetic superwind before it can be turned 
into stars.  \citet{pett00,pett02} have shown that cB58 is driving a 
wind in which two thirds of the $7 \times 10^{20}\,{\rm cm^{-2}}$ neutral 
column is blueshifted by $\sim 170\,{\rm km\,s^{-1}}$ with respect to $z_{\rm 
CO}$ (Table \ref{t-z}).  The black cores for the strongest absorption lines at 
this $z_{\rm abs}$ indicate a high covering fraction and suggest that much of 
the outflowing material lies well in front of the galaxy's stars along the 
line of sight.  Following \citet{heck00}, the mass loss rate for a 
constant--velocity, mass--conserving superwind flowing from some minimum 
radius $R_{\rm min}$ into a solid angle $\Delta \Omega_{\rm wind}$ will be
\begin{equation}\label{e-wind}
{\frac {{\dot M}_{\rm wind}}{M_\odot\,{\rm yr}^{-1}}} \simeq 
12\,\Big({\frac {R_{\rm min}}{\rm kpc}}\Big)\,\Big({\frac {v_{\rm wind}}
{170\,{\rm km\,s^{-1}}}}\Big)\,\Big({\frac {N_{\rm H}}{7 \times 10^{20}\,{\rm 
cm^{-2}}}}\Big)\,\Big({\frac {\Delta\Omega_{\rm wind}}{4\pi}}\Big)
\end{equation}
For cB58, setting $R_{\rm min} \simeq R_{\rm UV} = 2.0\,h_{0.7}^{-1}\,{\rm 
kpc}$ and $\Delta \Omega_{\rm wind}/4\pi \simeq 0.4 \pm 0.2$ by analogy with 
local systems \citep{heck00} yields ${\dot M}_{\rm wind} \simeq (10 \pm 
5)\,h_{0.7}^{-1}\,M_\odot\,{\rm yr}^{-1}$, somewhat smaller than the current 
star formation rate.  In this simple picture, the superwind should reduce 
$t_{\rm SF(future)}$ below $t_{\rm gas}$ by about a third-- an effect whose
relative mildness accords with several other lines of evidence.  Since $z_{\rm 
abs} \neq z_{\rm CO}$, the bulk of the molecular gas in cB58 cannot be 
participating in the outflow, and the observed enrichment of cB58 implies it 
has been able to retain at least some of the metals produced at earlier times 
\citep{pett02}.  Moreover, while $t_{\rm SF(future)} \sim t_{\rm gas}$ would 
already fall within a factor of 2--3 of cB58's most plausible $t_{\rm 
SF(past)} \sim 100 - 140\,{\rm Myr}$, a much lower $t_{\rm SF(future)} \ll 
t_{\rm SF(past)}$ would require that we have caught the starburst in this 
system at a special moment near its very end.

Overall, then, we appear to be observing a galaxy in the midst of converting 
$M_{\rm bary} \sim 10^{10}\,M_\odot$ worth of baryons into stars over a period 
$t_{\rm SF(past)} + t_{\rm SF(future)} \lesssim 420\,{\rm Myr}$ (see also \S 
\ref{ss-lbg} below).  Both the mass of stars formed (and their age) by the end 
of this process will fall within the ranges inferred from population synthesis 
analyses of large LBG samples \citep{papo01,shap01}.  As expected from 
considerations of causality \citep{lehn96}, the duration of the star formation 
episode will exceed the crossing time in cB58, which is roughly $t_{\rm cross} 
\simeq 2\,r_{\rm CO}\,D_{\rm A}/\sigma_{\rm CO} = 53\,h_{0.7}^{-1}\,{\rm 
Myr}$.  By the present day, cB58 will be of order a $0.07\,M_{\rm 
stars}^*$ system in the local mass function of \citet{cole01b},\footnote{For 
self--consistency, we assume here that $M_{\rm stars}^* = 1.4 \times 
10^{11}\,h_{0.7}^{-2}\,M_\odot$, as appropriate for the \citet{cole01b} 
population synthesis models using a $0.1-100\,M_\odot$ Salpeter IMF.} with the 
stars formed in the current burst ranging from $(10.9 - 
11.3)\,h_{0.7}^{-1}\,{\rm Gyr}$ in age, i.e., indistinguishable from coeval.  
This projected stellar mass is necessarily quite uncertain, given that the
superwind may permanently expel some fraction of the system's current baryonic 
mass, while galaxy merging, and ongoing star formation from the atomic gas 
presumably lying at larger radii, are both likely to occur as time passes.  
Nevertheless, it is suggestive that even this crude estimate lies well 
within the range of bulge masses estimated from $K$--band isophote fitting of 
local spirals.  The samples of Andredakis, Peletier, \& Balcells (1995) and 
Moriondo, Giovanardi, \& Hunt (1998), for example, scatter (widely) about a 
median $M_{\rm bulge} = 3-4 \times 10^{10}\,h_{0.7}^{-2}\,M_\odot$, assuming 
$(M/L)_{\rm bulge} \simeq 0.6\,M_\odot/L_\odot$ in $K$ band \citep{mori98}.

Because cB58 is such a strongly lensed system, it is fair to ask how 
our conclusions depend on the details of the lensing model.  The complexity of 
the S98 model, which is tightly constrained by a number of multiply 
imaged sources in the cluster field, makes a detailed examination prohibitive. 
However, we can at least argue that {\it differential} lensing of emission at 
different wavelengths is not likely to have undermined our results.  We have 
already assumed without penalty that ${\cal M}_{\rm CO} \simeq {\cal M}_{\rm 
UV}$ (in \S \ref{ss-cbpara}) and that ${\cal M}_{\rm UV} \simeq {\cal M}_{\rm 
H\alpha}$ (above).  The strong $z = 0$ correlation between dense gas and dust 
emission suggests ${\cal M}_{\rm FIR} \simeq {\cal M}_{\rm CO}$ is also 
reasonable.  More concretely, \citet{shap03} show that cB58 obeys all of the 
same scaling relations (among ${\rm Ly\,\alpha}$ equivalent width, 
interstellar absorption line strengths, and interstellar absorption 
line velocity offsets from ${\rm Ly\,\alpha}$) obeyed by unlensed LBGs.  This 
consistency suggests that the fraction ($f_{\rm arc} \sim 0.66$) of the 
background source we are seeing is not some unrepresentative slice of the 
galaxy.  In short, differential lensing does not appear to have a major impact 
on the inferred global parameters of cB58, or at least, is not the dominant 
source of uncertainty in our conclusions.  It remains possible, of course, to 
make assumptions that will lead to a contradiction of some sort.  Adopting a 
larger conversion factor $X_{\rm CO} \simeq 7.3 \times 10^{20}\,{\rm 
cm^{-2}\,(K\,km\,s^{-1})^{-1}}$ and a smaller age for the current star 
formation episode $t_{\rm SF(past)} = 20\,{\rm Myr}$, for example, the 
stellar/baryonic mass ratio becomes too small ($\sim 4\%$) to account for the 
observed enrichment.  However, for the equally (if not more) reasonable 
assumptions made in this paper, no such problems occur; this not only gives us 
confidence that the parameters we have derived are appropriate for cB58, but 
also hope that they may be representative for the LBG population as a whole.

\subsection{Implications for galaxy evolution models} \label{ss-lbg}

Our detection of CO(3--2) emission from cB58 provides a set of parameters 
whose values can be compared with those adopted and inferred by models for 
galaxy evolution at high redshift.  Basing such comparisons on rather 
uncertain measurements for a single object obviously limits the generality
of our conclusions; nevertheless, the exercise is interesting as a guide to 
the analyses that will eventually be possible for larger samples with more 
sensitive millimeter interferometers like ALMA.  Our discussion here focuses 
on comparisons to representative semi--analytic models (Kauffmann \& 
Haehnelt 2000 = KH00; Somerville, Faber, \& Primack 2001 = SPF01; Cole et al. 
2001a = CLBF01) rather than fully numerical models, since it is fairly 
straightforward to compare the assumptions made in the former with the 
relationships among the global properties we have measured for cB58.  The 
first question we examine is how well cB58 matches the models' 
parametrizations for star formation.  SPF01 assume (as in the fiducial ${\rm 
\Lambda CDM.3}$ model of Somerville \& Primack 1999) that for ``constant 
efficiency'' quiescent--mode star formation, $t_{\rm gas} = 50\,{\rm Myr}$ 
at all times.  This prescription overpredicts the current star formation rate 
in cB58 by about a factor of five.  Other authors have chosen to model 
quiescent--mode star formation using empirical relations seen to hold in local 
disk galaxy samples.  We have already noted in \S \ref{ss-cbsf} above that 
cB58's surface densities of (molecular) gas and star formation follow a global 
Schmidt law; this validates the simple assumption made for LBGs by Mo, Mao, \& 
White (1999).  Star formation rates also correlate with the ratio of gas 
surface density to a disk's dynamical timescale \citep{kenn98,wong02}.  For 
the convention $t_{\rm dyn} \equiv \pi R_{\rm disk}/v(R_{\rm disk})$, i.e., 
half of the orbital timescale at the disk's outer edge, this relation takes 
the form locally 
\begin{equation} \label{e-tdyn}
{\frac {t_{\rm gas}}{\rm Myr}} = 3.1 \times 10^4\,\Big({\frac {\alpha}{0.1}}
\Big)^{-1}\,\Big({\frac {R_{\rm disk}}{\rm kpc}}\Big)\,\Big({\frac {v(R_{\rm 
disk})}{\rm km\,s^{-1}}}\Big)^{-1}
\end{equation}
in terms of a dimensionless efficiency parameter $\alpha \simeq 0.1$ 
\citep{kenn98}.  For cB58, adopting $R_{\rm disk} \simeq 
2.0\,h_{0.7}^{-1}\,{\rm kpc}$ and $v\,(R_{\rm disk}) \simeq 74\,{\rm 
csc}\,i\,{\rm km\,s^{-1}}$ gives $t_{\rm dyn} \simeq 83\,{\rm sin}\,i\,
h_{0.7}^{-1}\,{\rm Myr}$ and $\alpha \simeq 0.3\,{\rm sin}\,i\,h_{0.7}^{-1}$ 
as observables for the models to reproduce.

In KH00, the disk dynamical timescale is estimated from the parameters 
of halos at a given redshift under the assumption that disks typically 
contract by a factor of ten before becoming rotationally supported.  In this 
case,
\begin{equation}\label{e-gamk}
{\frac {t_{\rm dyn}}{\rm Myr}} = 9.8 \times 10^3\,\Big({\frac {H(z)}{\rm 
km\,s^{-1}\,Mpc^{-1}}}\Big)^{-1}
\end{equation}
in terms of the Hubble constant as a function of redshift; for our cosmology, 
$H(2.729) = 282\,h_{0.7}\,{\rm km\,s^{-1}\,Mpc^{-1}}$ and $t_{\rm dyn} \simeq 
35\,h_{0.7}^{-1}\,{\rm Myr}$.  cB58 would then be expected to have 
$\alpha = t_{\rm dyn}/t_{\rm gas}\simeq 0.1\,h_{0.7}^{-1}$, compatible with 
the observations given its uncertain inclination.  However, $\alpha \simeq
0.1$ also represents almost a factor of four discrepancy with the 
parametrization of quiescent--mode star formation preferred by KH00, 
which postulates an efficiency declining as $\alpha(z) \propto (1+z)^{-1}$.  
A similar underefficiency is seen in the models of CLBF01, whose fiducial 
case effectively specifies that
\begin{equation} \label{e-dur}
\alpha = 0.005\,\pi\,\Big({\frac {v_{\rm disk}}{200\,{\rm km\,s^{-1}}}}
\Big)^{1.5}
\end{equation}
for the convention adopted in Equation \ref{e-tdyn} above and $v_{\rm 
disk}$ the circular velocity at the half--mass radius.  Assuming, crudely, 
that $v_{\rm disk} \simeq 74\,{\rm csc}\,i\,{\rm km\,s^{-1}}$ for cB58, 
Equation \ref{e-dur} predicts an efficiency $\alpha \simeq 0.0035\,{\rm 
csc}^{1.5} i$ that will only match the observations if $i \simeq 10^\circ$-- 
a scenario in which the galaxy's dynamical mass becomes implausibly large.

Although the above discussion suggests that current semi--analytic models do 
not adequately describe cB58 in terms of quiescent--mode star formation, it
remains possible that a merger--induced burst mode may provide a better 
account.  KH00 model bursts using a fixed gas consumption timescale 
$t_{\rm gas} = 100\,{\rm Myr}$; this is about a factor of three shorter 
than what we infer for cB58.  SPF01 implement a more complex prescription for 
bursts, in which a fraction $f_{\rm consume}$ of the progenitors' original gas 
mass is turned into stars according to
\begin{equation} \label{e-spf1}
{\rm SFR}\,(t) = {\rm SFR}_{\rm max}\,{\rm exp}\,\Big(-0.5\,(t/t_{\rm 
burst})^2\Big)
\end{equation}
over a duration $-2\,t_{\rm burst} \leq t \leq 2\,t_{\rm burst}$.  
Reflecting the results of numerical simulations, $f_{\rm consume}$ is set to 
0.75 for equal--mass mergers, and to 0.5 for minor (10:1 mass ratio) mergers 
in which neither progenitor has a significant bulge.  For cB58, if we fix 
$f_{\rm consume}$ and assume that the current stellar mass within $2.0\,
h_{0.7}^{-1}\,{\rm kpc}$ can be estimated independently of the galaxy's star 
formation history (e.g., as $M_{\rm stars} \simeq M_{\rm dyn} - M_{\rm gas}$), 
we can solve for ${\rm SFR}_{\rm max}$ and $t_{\rm burst}$ as well as the 
age of the current burst.  As a major merger in this parametrization, cB58  
would be observed $\simeq 14\,{\rm Myr}$ before the peak of a $t_{\rm burst} 
\simeq 130\,{\rm Myr}$ burst; as a minor merger, it would be observed $\simeq 
35\,{\rm Myr}$ after the peak of a $t_{\rm burst} \simeq 79\,{\rm Myr}$
burst.  SPF01 set $t_{\rm burst} \propto t_{\rm dyn}$ in their models; the
$t_{\rm dyn} \simeq 83\,{\rm sin}\,i\,h_{0.7}^{-1}\,{\rm Myr}$ estimated for
cB58 above thus slightly favors a minor merger scenario, in which the bulge 
that could otherwise have suppressed star formation \citep{miho94} might 
only now be forming (see \S \ref{ss-cbsf}).  However, since SPF01 also assume 
that the quiescent mode of star formation should continue throughout a burst, 
the total star formation rate their model predicts for cB58 is still likely to 
exceed what is observed.

In the context of the semi--analytic models considered here, then, cB58's star 
formation parameters can be accounted for by a simple Schmidt law; by a 
quiescent mode whose rate depends on the dynamical timescale (provided 
the efficiency $\alpha$ remains at least as high as seen locally; cf. KH00); 
or by a burst mode triggered by a minor merger (provided rapid quiescent--mode 
star formation can somehow be suppressed; cf. SPF01).  Given the observational 
uncertainties, the variety of complex star formation histories that are 
compatible with cB58's rest--UV spectrum (e.g., Leitherer et al. 2001), and 
the lack of comparison objects, it would be premature for us to argue strongly 
here for or against the use of any of these star formation recipes for the 
LBG population as a whole.  cB58 does fulfill the more general expectation-- 
noted by both KH00 and SPF01-- that star--forming galaxies at high redshift 
should be relatively gas--rich.  SPF01 in particular predict that LBGs should 
have high mass fractions of cold (molecular) gas, with the median $f_{\rm gas} 
\sim 0.5$ for LBGs with $V_{606} < 25.5$.  For cB58, whose observed $V$ 
magnitude \citep{elli96} corrected for lensing implies $V \simeq 24.4$, 
the most likely $f_{\rm gas} \sim 0.66$ falls well inside the envelope of 
the authors' mock Hubble Deep Field catalogue (as is also the case for several 
of the galaxy's other parameters; see their Figure 16).  The KH00 models 
predict similarly that $f_{\rm gas}$ for major mergers increases to $\sim 0.6$ 
by $z \sim 2.7$, although the evolution will not be as strong for a ${\rm 
\Lambda CDM}$ cosmology.

A second question our observations of cB58 help address relates to the 
uncertain mass scale of LBGs.  Direct estimates of stellar masses from 
optical/near--IR photometry \citep{papo01,shap01} and dynamical masses from 
nebular emission line widths \citep{pett98,pett01} reveal $M_{\rm bary} \sim 
10^{10}\,M_\odot$ concentrations in the centers of LBGs, as anticipated by 
early analyses of LBG sizes and luminosities alone (e.g., Lowenthal et al. 
1997).  In this context, it is not a surprise that we estimate $M_{\rm bary} 
\sim 10^{10}\,M_\odot$ in cB58 from observations that (by assumption-- most 
explicitly in our invocation of ${\cal M}_{\rm CO} \simeq {\cal M}_{\rm UV}$) 
again trace only a galaxy's bright core.  The fact that the CO(3--2) linewidth 
is within 10\% of the ${\rm H\alpha}$ linewidth measured by \citet{tepl00}-- 
$\sigma_{\rm CO} = 74\,{\rm km\,s^{-1}}$ vs. $\sigma_{\rm H\alpha} = 81\,{\rm 
km\,s^{-1}}$-- does supply useful evidence that ${\rm H\alpha}$ is not unduly 
affected by the superwind, although it remains possible that neither 
${\rm H\alpha}$ nor CO emission traces cB58's full rotation curve.  The more 
unique contribution here comes from our estimate of a total star formation 
timescale ($t_{\rm SF(past)} + t_{\rm SF(future)} \lesssim 420\,{\rm Myr}$ for 
a constant star formation rate; $4 \times t_{\rm burst} \sim 320\,{\rm Myr}$ 
for a merger--triggered burst a la SPF01) that is a substantial fraction of 
the time elapsed over the $2.5 \leq z \leq 3.5$ range probed by the Lyman 
break selection function ($t_{\Delta z} \sim 800\,{\rm Myr}$ for our 
cosmology).  Given that the star formation timescale can be extended if the 
gas reservoir is replenished by inflow, recycling (Kennicutt, Tamblyn, \& 
Congdon 1994), or future merging, we can conclude-- provided cB58 is 
representative and does not suffer substantial mass loss-- that the duty 
cycle for star formation in LBGs at this redshift will be of order 50\%. 

The strong clustering of LBGs implies that they are a highly biased 
galaxy population \citep{giav98,adel98,arno99,giav01,porc02,fouc03}, 
and thus that they must occupy the most massive halos at $z \sim 3$.  Exactly 
how to associate LBGs with halos of mass $M$-- i.e., the precise form of the 
halo occupation function $N_{\rm g}(M)$-- remains an open issue.  If $N_{\rm 
g} = 1$ is assumed, a fiducial $M$ can be estimated for a given LBG sample by 
identifying the mass scale at which halos have the same number density and 
bias.  \citet{adel98} and \citet{giav01} use this approach to infer $M \sim 
1.1 \times 10^{12}\,h_{0.7}^{-1}$ for the halos hosting LBGs with ${\cal R} 
\leq 25.5$ (like cB58 after correction for lensing).  The associated baryonic 
masses should then be $M_{\rm bary} \sim 1.5 \times 10^{11}\,
M_\odot$,\footnote{We assume $M/M_{\rm bary} = \Omega_{\rm M}/\Omega_{\rm b}$ 
for $\Omega_{\rm b} = 0.04\,h_{0.7}^{-2}$ \citep{burl98}.} which even if 
reduced to match lower estimated clustering strengths \citep{adel03} will be 
in excess of the direct measurements for most individual LBGs by at least 
a factor of a few. Packing multiple LBGs into a single massive halo would in 
principle reduce this discrepancy.  However, while the high close pair 
fraction at $z \sim 3$ does favor $dN_{\rm g}(M)/dM > 0$ (Wechsler et al. 
2001; Bullock, Wechsler, \& Somerville 2002), the apparent dependence of 
clustering strength on number density \citep{giav01,fouc03} argues strongly 
that in fact $dN_{\rm g}(M)/dM \leq 0$ \citep{bull02}.  Our estimate of a 
$\sim 50\%$ duty cycle for star formation further suggests that the 
remaining baryons in a halo hosting one LBG also do not lie in some large 
population of cohabiting, ``dormant'' LBGs in which starbursts have faded or 
not yet begun.  This consideration further disfavors scenarios in which star 
formation episodes in an individual LBG are extremely brief (e.g., triggered 
by frequent sub--halo collisions: Kolatt et al. 1999).  Instead, the remaining 
baryons in a given LBG's host halo are likely to be found in the galaxy's own 
lower surface brightness outskirts, in cohabiting galaxies of significantly 
lower mass, or outside galaxies altogether (where they may have been 
banished by a vigorous superwind).

\subsection{Millimeter line/continuum ratios at $z \sim 3$} \label{ss-ltoc}

After SMM\,J14011+0252 \citep{fray99,ivis00}, cB58 is only the second galaxy 
without an active nucleus to have been detected in CO emission at high 
redshift.  Among all detected sources, cB58 is also fairly unusual in not 
having been prioritized for CO observations on the basis of a previous 
measurement of strong dust emission in the (sub)millimeter; the radio galaxy 
53W002 \citep{scov97} is perhaps the only other member of this category.  
Because it does not reflect common selection biases, then, it is of interest 
to compare cB58's line/continuum ratio to those of other systems at comparable 
redshifts.  Due to the large uncertainties in dust temperature and $X_{\rm 
CO}$ conversion factor, this ratio may only poorly translate to a star 
formation efficiency or a gas/dust ratio.  
However, it is still interesting from a practical standpoint to see how much 
source--to--source variation in global ISM parameters exists at high redshift, 
since this will drive strategies for deep surveys with the next generation of 
millimeter interferometers.  

In order to limit the impact of CO excitation uncertainties on our comparison, 
we will restrict our comparison to sources in the range $2.3 \leq z 
\leq 2.8$ that have been observed in the CO(3--2) line and in continuum at 
1.2\,mm.  The observed 1.2\,mm flux density (corresponding to rest wavelengths 
$316 - 363\,{\rm \mu m}$ over this redshift range) is a generally reliable 
(albeit somewhat redshift--dependent) measure of thermal dust 
emission.\footnote{We exclude the exception to this rule, MG\,0414--0534, 
whose extremely strong 1.2\,mm flux density is dominated by nonthermal 
emission \citep{barv98}.}  cB58 has a ratio ${\cal R}_{\rm L/C} \equiv F_{\rm 
CO}/S_{1.2} \simeq 350\,{\rm km\,s^{-1}}$; if we assume as discussed in \S 
\ref{ss-ximp} that 0.22\,mJy of its apparent 1.2\,mm flux density actually 
comes from the foreground cD, then ${\cal R}_{\rm L/C}$ becomes $\sim 440\,
{\rm km\,s^{-1}}$.  The seven comparison objects in the compilation of 
\citet{beel03}\footnote{F10214+4724, the Cloverleaf, 53W002, 
SMM\,J02399--0136, Q1230+1627B, SMM\,J14011+0252, and J\,1408+5628.} have 
values of ${\cal R}_{\rm L/C}$ ranging from $300\,{\rm km\,s^{-1}}$ 
(J\,1408+5628: Omont et al. 2003; Beelen et al. 2003) at the low end to 
$700\,{\rm km\,s^{-1}}$ (53W002: Alloin, Barvainis, \& Guilloteau 2000) and 
$800\,{\rm km\,s^{-1}}$ (SMM\,J14011+0252) at the high end.  With cB58 
occupying the opposite end of this interval from the only other galaxy not 
hosting an energetically dominant active nucleus (SMM\,J14011+0252), and the 
only other galaxy not selected on the basis of its dust emission (53W002), 
there is no evidence that either selection criterion significantly biases the 
line/continuum ratio.  It is in fact rather remarkable that all eight systems 
have ${\cal R}_{\rm L/C}$ values agreeing to within better than a factor of 
three.  Two quasars with CO(3--2) upper limits in the survey of \citet{guil99} 
would in principle correspond to ${\cal R}_{\rm L/C} < 300\,{\rm km\,s^{-1}}$; 
however, since quasar nondetections often result from uncertain CO redshifts 
rather than weak CO lines, these may not be the outliers they appear.  

\section{Conclusions} \label{s-conc}

From our observations of the gravitationally lensed Lyman break galaxy cB58, 
we estimate that within a radius of $2.0\,h_{0.7}^{-1}\,{\rm kpc}$, it most 
likely has
\begin{enumerate}

\item an age for the current star formation episode $t_{\rm SF(past)} \sim 
140\,{\rm Myr}$;

\item a current star formation rate $= (23.9 \pm 0.7)\,
h_{0.7}^{-2}\,M_\odot\,{\rm yr^{-1}}$;

\item a formed stellar mass $M_{\rm stars} \sim 3.4 \times 10^9\,
h_{0.7}^{-2}\,M_\odot$;

\item a molecular gas mass $M_{\rm gas} = 6.6^{+5.8}_{-4.3} \times 
10^9\,h_{0.7}^{-2}\,M_\odot$;

\item a lifetime for continued star formation $t_{\rm SF(future)} \leq t_{\rm 
gas} = 280^{+280}_{-180}\,{\rm Myr}$, lower to the extent that molecular 
gas is expelled by the observed superwind before it can be turned into stars;

\item a molecular gas mass surface density $\Sigma_{\rm gas} = 
540^{+470}_{-350}\,M_\odot\,{\rm pc^{-2}}$;

\item a dynamical mass $M_{\rm dyn} = 1.0^{+0.6}_{-0.4} \times 
10^{10}\,{\rm csc}^2 i\,h_{0.7}^{-1}\,M_\odot$; and

\item a gas mass fraction $f_{\rm gas} = 0.66^{+0.34}_{-0.52}\,{\rm 
sin}^2 i\,h_{0.7}^{-1}$.

\end{enumerate}
We judge that these estimates should not have been severely compromised by 
differential lensing, or by the possible effects of nonthermal continuum or 
(lensed) CO line emission from a nearby source X.  We conclude that cB58 is 
partway through an episode of star formation lasting several dynamical 
timescales, and in the absence of substantial mass loss, gas inflow, or 
further merging will evolve into a $z = 0$ system whose mass and coevality are 
those of a spiral bulge.  

To the extent that cB58 is a typical LBG apart from its magnification, its 
global parameters place new constraints on the permissible recipes for star 
formation in semi--analytic models of galaxy evolution.  Its current star 
formation rate, molecular gas mass, and inferred stellar mass are at least 
reasonably compatible with the expectations for a Schmidt law, for a quiescent 
mode of star formation tied to the local dynamical timescale, and for a burst 
mode of star formation (here triggered by a minor merger), although no 
published model offers perfect agreement.  Its high gas mass fraction agrees 
with theoretical predictions that $f_{\rm gas}$ should increase with redshift.
The duration of cB58's current star formation episode appears to be long 
relative to the redshift range probed by the Lyman break technique at $z \sim 
3$; this argues for a high duty cycle for star formation, and against 
scenarios in which the halo hosting an observed LBG is over time ``lit up'' by 
many different objects undergoing brief bursts.  Finally, from a comparison 
with other $z \sim 2.5$ systems detected in CO(3--2) and dust emission, we 
arrive at the useful empirical conclusion that the millimeter line/continuum 
ratio is not a strong function of nuclear activity or dust luminosity.

\acknowledgments

We are grateful to IRAM staff members at the PdBI and in Grenoble, 
particularly Raphael Moreno and Roberto Neri, for their assistance in 
conducting the observations and advice on reducing the data.  We thank Stella 
Seitz for providing reduced {\it HST} images, as well as for her insights on 
lens models, and acknowledge helpful comments from an anonymous referee 
and useful discussions with Andrea Cimatti, Pierre Cox, Bruce Elmegreen, 
David Hogg, Guinevere Kauffmann, Masami Ouchi, Max Pettini, Alice Shapley, 
Rachel Somerville, and Harry Teplitz.  This work has made use of the 
NASA/IPAC Extragalactic Database (NED), which is operated by the Jet 
Propulsion Laboratory, California Institute of Technology, under contract with 
NASA.

\appendix

\section{Revisiting the S98 lensing model} \label{a-lens}

\begin{figure}
\plotone{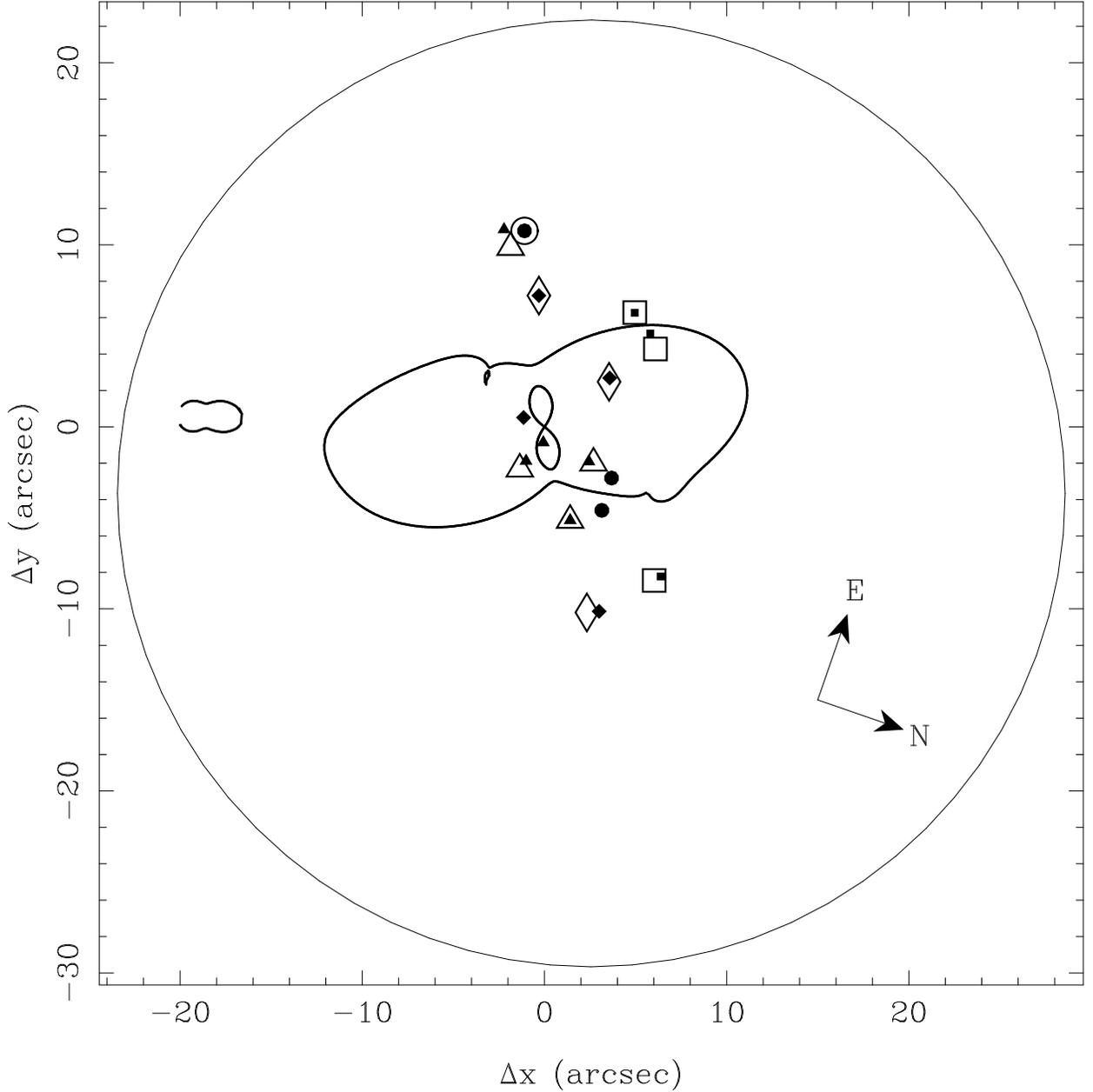}
\caption{The S98 lensing model as implemented in {\it gravlens}, with 
coordinate offsets relative to the location of the cD and parallel to the 
rows and columns of the WFPC2 chip.  For each image set (A = circles, B = 
squares, C = diamonds, W = triangles), open symbols show observed images and 
solid symbols show predicted images.  One model image in each set (A2, B1, C1, 
and W1) has been forced to agree with its observed location; the reasonable 
agreement between model and data for the {\it other} images of the 
corresponding background source verifies that the model works reasonably well. 
The thick lines show the image--plane critical curves for a background source 
at $z \simeq 2.73$ (compare e.g., Figure 14 of S98), only directly relevant 
for the A image set that includes cB58; the thin circle shows the 50\% 
response of the PdBI primary beam at 92.96\,GHz, whose diameter is $\sim 
52\arcsec$.
\label{f-s98}}
\end{figure}

S98 model the total deflection potential of the lens as a sum of 
elliptical potentials due to individual galaxies and the cluster as a whole.
The $i$th term in this sum is centered at some position $(x_i,y_i)$ and is 
characterized by a normalization $\Theta_{{\rm E},i}$ (recoverable from a 
tabulated velocity dispersion $\sigma_i$ according to their equation A3), a 
scale length $\zeta_i$, and an ellipticity $\epsilon_i$:
\begin{equation}\label{e-spiep}
\phi_i\,(x,y) = \Theta_{{\rm E},i}\,\Big(\zeta_i^2 + {\frac {1-\epsilon_i}{1+ 
\epsilon_i}}\,(x - x_i)^2 + {\frac {1 + \epsilon_i}{1 - \epsilon_i}}\,(y - 
y_i)^2\Big)^{\frac 1 2}
\end{equation}
In {\it gravlens}, a softened power--law potential can be specified 
in terms of an index $\alpha_i$, normalization $b_i$, scale length $s_i$, and 
ellipticity $e_i$:
\begin{equation}\label{e-kpiep}
\phi_i\,(x,y) = b_i\,\Big(s_i^2 + (x - x_i)^2 + {\frac 1 {(1-e_i)^2}}\,(y - 
y_i)^2\Big)^{{\frac 1 2}\alpha_i} 
\end{equation}
Equality between Equations \ref{e-spiep} and \ref{e-kpiep} holds for $\alpha_i 
= 1$ and 
\begin{eqnarray}
b_i & = & \Theta_{{\rm E},i}\,\Big({\frac {1-\epsilon_i}{1+\epsilon_i}}
\Big)^{\frac 1 2} \label{e-bscale} \\
s_i & = & \zeta_i\,\Big({\frac {1+\epsilon_i}{1-\epsilon_i}}\Big)^{\frac 1 2}\\
e_i & = & {\frac {2\epsilon_i}{1 + \epsilon_i}} \label{e-escale} 
\end{eqnarray}
Our version of the S98 model is constructed in an $(x,y)$ coordinate frame 
with axes parallel to the rows and columns of the WFPC2 chip and an origin at 
the position of the cluster cD.  We have measured lens--plane galaxy positions 
$\{x_i,y_i\}$ from the F675W image of S98, used Equations \ref{e-bscale} 
through \ref{e-escale} to rescale the $\{\Theta_{{\rm E},i},\zeta_i,
\epsilon_i\}$ values specified for their model MVIa, and rotated position 
angles for the cluster and cD major axes to match our coordinate convention.  
To confirm that the resulting {\it gravlens} version is a reasonable facsimile 
of the original model, we have examined the lensing of a background point 
source which would appear at the observed position of image A2.  Our model 
specifies that the image of a point source at A2 has undergone a factor of 
2.38 magnification, in excellent agreement with the factor of 2.39 estimated 
by S98.  Our model also predicts that two additional images of the same 
background source should appear at the locations marked by filled circles in 
Figure \ref{f-s98}, i.e., within the cB58 arc.  Exact reproduction of the arc 
produced by an intrinsically extended (rather than point) background source is 
beyond the scope of this paper.

We have also checked that the projected mass density specified by our 
reparametrization of the S98 lensing model can roughly account for the 
positions and magnifications of the additional (B, C, and W) sets of multiple 
images in this field.  Because the corresponding background sources do not 
necessarily lie at the same $z \simeq 2.73$ as cB58, it is necessary to apply 
a further rescaling to the deflection potential for each one.  Conveniently, 
the cosmological dependence of the deflection potential can be factored out 
of the normalizations for the individual terms.  For source and lens redshifts 
$z_{\rm s}$ and $z_{\rm l}$ and the critical surface density for lensing 
$\Sigma_{\rm crit}$,
\begin{equation}
\phi_i\,(x,y) \propto \Sigma_{\rm cr}^{-1} = {\frac {4\pi G}{c^2}}\,{\frac 
{D_{\rm A}\,(z_{\rm l},0)\,D_{\rm A}\,(z_{\rm s},0)}{D_{\rm A}\,(z_{\rm s},
z_{\rm l})}}
\end{equation}
where we write $D_{\rm A}\,(z_2,z_1)$ to denote the angular diameter distance 
to a source at redshift $z_2$ as seen from a lower redshift $z_1$.  The 
correct rescaling for the $j$th background source at redshift $z_j$ is thus to 
map $b_i \rightarrow b_i\,d_j$ for each $i$, where $d_j$ is a lensing strength 
defined as
\begin{equation}
d_j\,(z_j,\Omega_M,\Omega_\Lambda) = {\frac {D_{\rm A}\,(z_j,0.37)}
{D_{\rm A}\,(z_j,0)}}\,\Big[{\frac {D_{\rm A}\,(2.73,0.37)}
{D_{\rm A}\,(2.73,0)}}\Big]^{-1}
\end{equation}
The precise mapping between $d_j$ and $z_j$ depends on the choice of cosmology.
Since the S98 model MVIa solves for the $\{d_j\}$ explicitly (corresponding, 
in our cosmology, to $z_B \simeq 3.05$, $z_C \simeq 3.57$, and $z_W \simeq 
2.85$), we can rescale our {\it gravlens} deflection potential directly.  The 
derived image positions shown in Figure \ref{f-s98} are in generally good 
agreement with the data, although {\it gravlens} predicts that there should be 
one additional image each of both the C and the W background sources (only the
latter is predicted by S98).  Note that the assumption of $z_X = 1.48$ (see 
below) fixes $d_X = 0.874$ in our cosmology.

\begin{figure}
\includegraphics{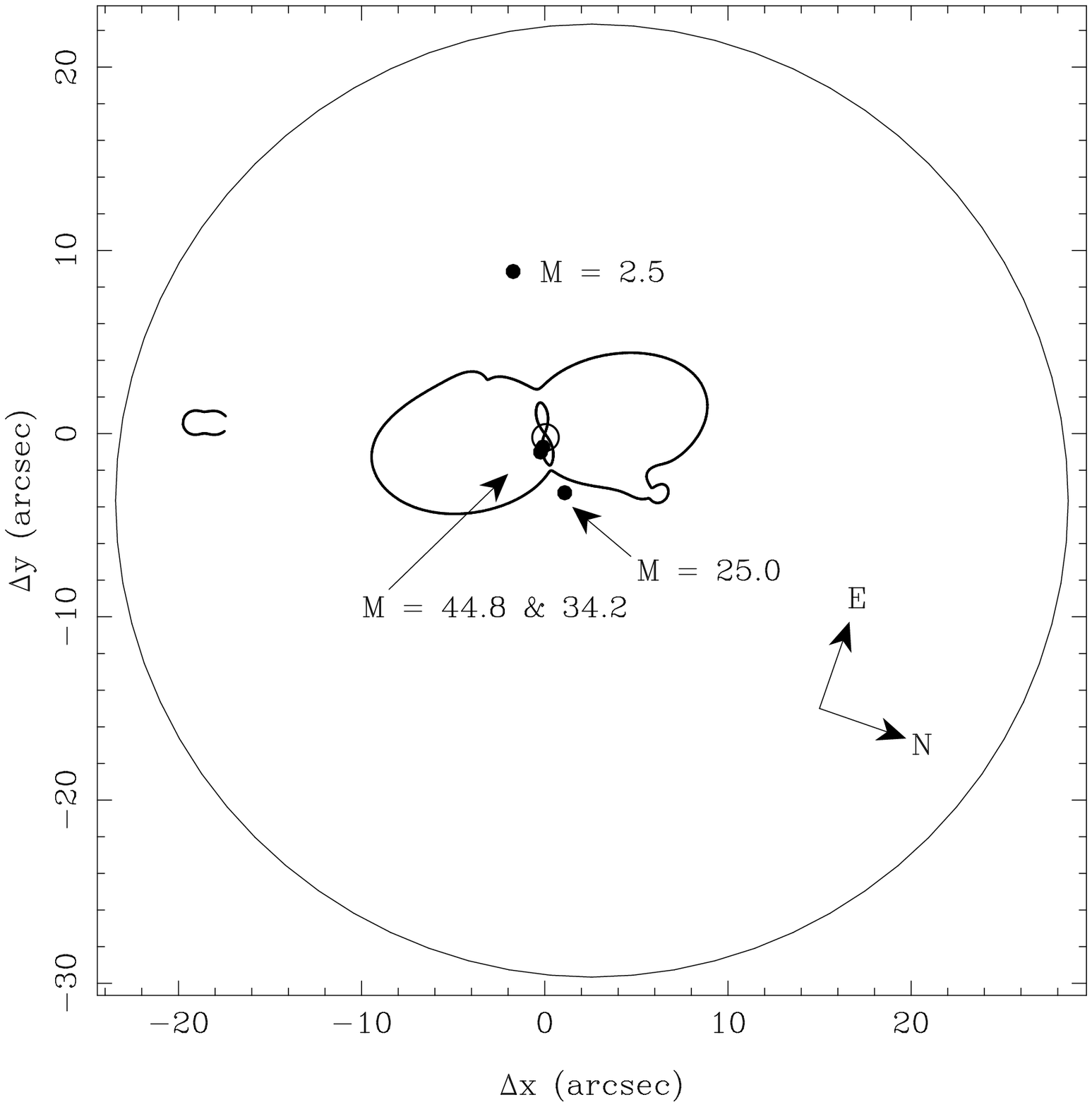}
\includegraphics{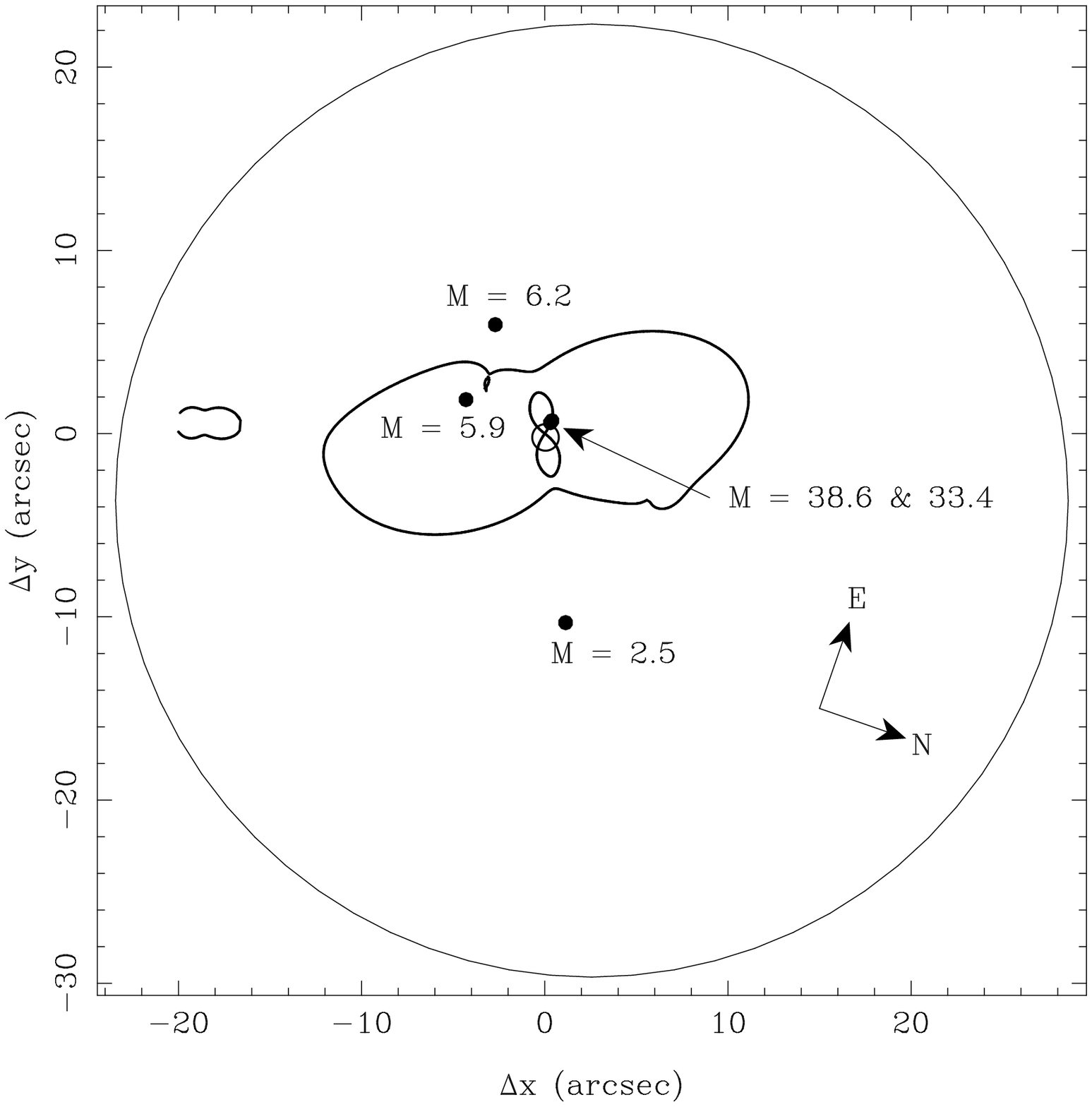}
\includegraphics{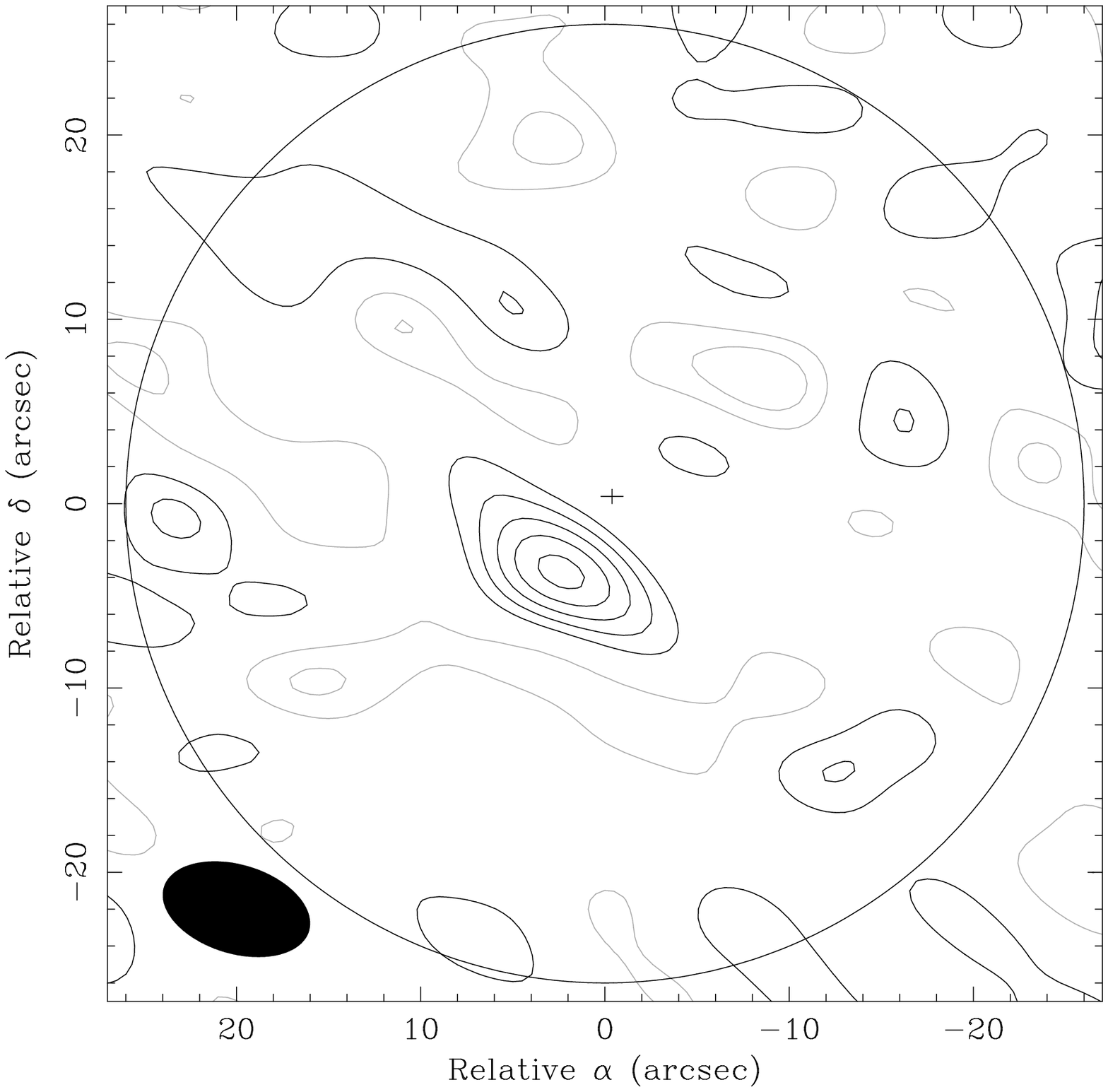}
\vspace{7.0in}
\caption{Lower panel: same as center panel of Figure \ref{f-xmap}, but with a 
larger field of view and contours that are multiples of $1\sigma$ ($0.16\,{\rm 
mJy\,beam^{-1}}$).  Upper panels: possible lens models for $z_X \simeq 1.48$ 
(left) and $z_X \simeq 2.73$ (right), in coordinates defined relative to the 
cD position and rotated into alignment with the WFPC2 chip.  Small circles at 
origin show the nominal position of X; filled circles show the positions of 
predicted images and are labelled with magnification factors.  The thick lines 
show the image--plane critical curves for a source at $z_X$.  The thin circle 
in all three panels shows the 50\% response of the PdBI primary beam at 
92.96\,GHz.
\label{f-2lens}}
\end{figure}

In exploring the possibility that source X could be gravitationally lensed 
CO(2--1) emission from a source at $z_X \simeq 1.48$ or CO(3--2) emission
from a source (other than cB58) at $z_X \simeq 2.73$ (see \S \ref{ss-xor} 
above), the strongest constraint is 
imposed by the absence of counterimages within the 50\% response of the 
PdBI primary beam at 92.96\,GHz (see, e.g., the lower panel of Figure 
\ref{f-2lens}).  Indeed, the second--strongest feature in the entire map is a 
$4\sigma$ peak lying $\sim 57\arcsec$ to the southeast of X (not shown), where 
the primary beam response is only $\sim 3\%$.  If this were a true 
counterimage, its implied flux density of $20.4 \pm 5.0\,(\pm 3.0)\,{\rm mJy}$ 
would require that it be more highly magnified than X by a factor between 15 
and 36-- an unlikely situation at a position where the lens model predicts 
only a single image and a magnification $\sim 1.1$ for either choice of $z_X$. 
The counterimages of X that the model does predict lie at much smaller radii.  
For both $z_X \simeq 1.48$ and $z_X \simeq 2.73$, we have used {\it gravlens} 
to predict the image positions and magnifications that result if we place X at 
each of the nine grid points defined by $(\alpha \pm \Delta \alpha, \delta \pm 
\Delta \delta)$ in terms of the astrometric uncertainties listed in Table 
\ref{t-pos}.  Typically, we find X has one or more counterimages that are at 
least as strongly magnified, far enough away not to be confused with X, and 
close enough to our phase center that they would hardly suffer any primary 
beam attenuation.  Such counterimages would be detected if they existed at 
all; their absence instantly invalidates the particular models in which they 
appear.  

In a few cases, the lensing geometry can conspire to place X and a bright 
counterimage X2 close to each other on opposite sides of a critical curve.  
The upper panels of 
Figure \ref{f-2lens} show the most promising examples of such configurations.  
For $z_X \simeq 1.48$ (at left), placing X at $(\alpha - \Delta \alpha, 
\delta)$ yields a net magnification of X+X2 three times higher than that of 
the next brightest counterimage in the field.  For $z_X \simeq 2.73$ (at 
right), the magnification of X+X2 that results from placing X at $(\alpha + 
\Delta \alpha,\delta)$ is a factor of twelve higher than that of the next 
brightest counterimage-- entirely consistent with the manifestation of only a 
``single'' source in the lower panel of Figure \ref{f-2lens}.  Exactly how 
realistic such models are depends strongly on the precise shape of the lens 
mass profile at small radius.  With \citet{kass93} having shown that 
elliptical potentials translate to unphysical, peanut--shaped isodensity 
contours for ellipticities $e > 0.18$ (using the {\it gravlens} definition), 
for example, our use of $e = 0.28$ for the cluster and $e = 0.23$ for the cD 
suggests there are reasonable grounds for caution.

\clearpage

\begin{deluxetable}{cclllll}
\tablecaption{Positions and positional uncertainties.
\label{t-pos}}
\tablewidth{0pt}
\tablehead{
\colhead{} & 
\colhead{$\lambda_{\rm obs}$} &
\colhead{$\alpha$} & 
\colhead{$\Delta \alpha$} & 
\colhead{$\delta$} & 
\colhead{$\Delta \delta$} &
\colhead{} \\
\colhead{Source} & 
\colhead{(observed)} &
\colhead{(J2000)} & 
\colhead{(J2000)} & 
\colhead{(J2000)} & 
\colhead{(J2000)} & 
\colhead{Reference(s)}
}
\startdata
phase center & --- & 15:14:22.25 & --- & +36:36:24.40 & --- & --- \\ 
\tableline
cB58${}^a$ & CO(3--2) & 15:14:22.218 & $\pm 0.050$ & +36:36:24.79 & $\pm 0.39$ 
& 1 \\
     & optical & 15:14:22.271${}^e$ & $\pm 0.010$ & +36:36:25.2${}^e$ & $\pm 
0.12$ & 2, 3 \\
\tableline
cD${}^b$ & 6.1\,cm & 15:14:22.47  & $\pm 0.002$ & +36:36:20.8  & $\pm 0.02$ & 
2 \\
         & 21\,cm  & 15:14:22.443 & $\pm 0.075$ & +36:36:20.68 & $\pm 0.90$ & 
4 \\
\tableline
S${}^c$ & optical & 15:14:22.578${}^e$ & $\pm 0.010$ & +36:36:16.8${}^e$ & 
$\pm 0.12$ & 2, 3 \\
\tableline
X & 3.2\,mm & 15:14:22.455 & $\pm 0.049$ & +36:36:20.90 & $\pm 0.38$ & 1 \\
\enddata
\tablenotetext{a}{cB58 = MS\,1512.4+3647:\,PPP\,101120 in \citet{abra98}.}
\tablenotetext{b}{cD = MS\,1512.4+3647:\,PPP\,101094 in \citet{abra98}.}
\tablenotetext{c}{S (S98) = MS\,1512.4+3647:\,PPP\,101068 in 
\citet{abra98}.}
\tablenotetext{e}{Calculated by identifying the \citet{laur97} 6\,cm radio 
core with the cD, and using the \citet{abra98} offsets from the optical cD 
position.}
\tablerefs{(1) this paper (2) \citet{laur97} (3) \citet{abra98} (4) 
\citet{beck95}}
\end{deluxetable}

\begin{deluxetable}{cclrl}
\tablecaption{Measured redshifts for cB58, from red to blue.
\label{t-z}}
\tablewidth{0pt}
\tablehead{
\colhead{} & 
\colhead{} & 
\colhead{} &
\colhead{Velocity${}^a$} & 
\colhead{} \\ 
\colhead{Quantity} & 
\colhead{Rest feature(s)} & 
\colhead{Redshift} &
\colhead{(${\rm km\,s^{-1}}$)} & 
\colhead{Reference}
}
\startdata
$z_{\rm H\,II}$ & mean of 10 ($\lambda\lambda\,3726 - 6583$) & $2.7290 \pm 0.0007$ & 0 & 1 \\
$z_{\rm stars}^b$  & \ion{C}{3} $\lambda\,2297$ & $2.7276^{+0.0001}_{-0.0003}$ & --110 & 2\\
$z_{\rm stars}^b$  & mean of 10 ($\lambda\lambda\,1294 - 1718$) & $2.7268 \pm 0.0008$ & --180 & 3 \\
$z_{\rm CO}$    & CO(3--2)             & $2.7265^{+0.0004}_{-0.0005}$ & --200 & 4 \\
$z_{\rm abs}^c$  & mean of 35 ($\lambda\lambda\,1134 - 2577$) & $2.7244$ & --370 & 2 \\
\enddata
\tablenotetext{a}{Velocity relative to $z_{\rm H\,II}$ in the source rest 
frame.}
\tablenotetext{b}{We list both reported estimates of $z_{\rm stars}$, 
although \citet{pett02} express a preference for $z_{\rm stars} = 2.7276$.}
\tablenotetext{c}{Redshift of deepest interstellar absorption.  \citet{pett02}
note that absorption at some level occurs over a velocity range of $1000\,{\rm 
km\,s^{-1}}$.}
\tablerefs{(1) \citet{tepl00} (2) \citet{pett02} (3) \citet{pett00} (4) this 
paper}
\end{deluxetable}

\clearpage

\clearpage

\clearpage

\clearpage

\clearpage

\clearpage

\begin{thebibliography}{}

\bibitem[Abraham et al.(1998)]{abra98}
  Abraham, R. G., Yee, H. K. C., Ellingson, E., Carlberg, R. G., \& Gravel, 
  P. 1998, \apjs, 116, 231

\bibitem[Adelberger \& Steidel(2000)]{adel00}
  Adelberger, K. L., \& Steidel, C.~C. 2000, \apj, 544, 218

\bibitem[Adelberger et al.(1998)]{adel98}
  Adelberger, K. L., Steidel, C. C., Giavalisco, M., Dickinson, M., Pettini, 
  M., \& Kellogg, M. 1998, \apj, 505, 18

\bibitem[Adelberger et al.(2003)]{adel03}
  Adelberger, K. L., Steidel, C. C., Shapley, A. E., \& Pettini, M. 2003, 
  \apj, 584, 45

\bibitem[Alloin, Barvainis, \& Guilloteau(2000)]{allo00}
  Alloin, D., Barvainis, D., \& Guilloteau, S. 2000, \apj, 528, L81

\bibitem[Andredakis et al.(1995)]{andr95}
  Andredakis, Y. C., Peletier, R. F., \& Balcells, M. 1995, \mnras, 275, 874

\bibitem[Arimoto et al.(1996)]{arim96}
  Arimoto, N., Sofue, Y., \& Tsujimoto, T. 1996, \pasj, 48, 275

\bibitem[Arnouts et al.(1999)]{arno99}
  Arnouts, S., Cristiani, S., Moscardini, L., Matarese, S., Lucchin, F., 
  Fontana, A., \& Giallongo, E. 1999, \mnras, 310, 540

\bibitem[Baker et al.(2001)]{bake01}
  Baker, A. J., Lutz, D., Genzel, R., Tacconi, L. J., \& Lehnert, M. D. 2001, 
  \aap, 372, L37

\bibitem[Barvainis et al.(1998)]{barv98}
  Barvainis, R., Alloin, D., Guilloteau, S., \& Antonucci, R. 1998, \apj, 492, 
  L13

\bibitem[Bechtold et al.(1998)]{bech98}
  Bechtold, J., Elston, R., Yee, H. K. C., Ellingson, E., \& Cutri, R. M. 
  1998, in ASP Conf. Ser. 146, The Young Universe, ed. S. D'Odorico, A. 
  Fontana, \& E. Giallongo (San Francisco: ASP), 241

\bibitem[Bechtold et al.(1997)]{bech97}
  Bechtold, J., Yee, H. K. C., Elston, R., \& Ellingson, E. 1997, \apj, 477, 
  L29

\bibitem[Becker et al.(1995)]{beck95}
  Becker, R. H., White, R. L., \& Helfand, D. J. 1995, \apj, 450, 559

\bibitem[Beelen et al.(2003)]{beel03}
  Beelen, A., et al. 2003, \aap, in preparation

\bibitem[Bolatto et al.(2003)]{bola03}
  Bolatto, A. D., Leroy, A., Israel, F. P., \& Jackson, J. M. 2003, \apj, 595, 
  167

\bibitem[Bullock et al.(2002)]{bull02}
  Bullock, J. S., Wechsler, R. H., \& Somerville, R. S. 2002, \mnras, 329, 246

\bibitem[Bryant \& Scoville(1999)]{brya99}
  Bryant, P. M. \& Scoville, N. Z. 1999, \aj, 117, 2632

\bibitem[Burles \& Tytler(1998)]{burl98}
  Burles, S. \& Tytler, D. 1998, \apj, 507, 732

\bibitem[Chapman et al.(2000)]{chap00}
  Chapman, S. C., et al. 2000, \mnras, 319, 318
% second author is Scott

\bibitem[Chapman et al.(2002)]{chap02}
  Chapman, S. C., Shapley, A., Steidel, C., \& Windhorst, R. 2002, \apj, 572, 
  L1

\bibitem[Cole et al.(2001a)]{cole01a}
  Cole, S., Lacey, C. G., Baugh, C. M., \& Frenk, C. S. 2001a, \mnras, 319, 
  168 (CLBF01)

\bibitem[Cole et al.(2001b)]{cole01b}
  Cole, S., et al. 2001b, \mnras, 326, 255
% second author is Norberg

\bibitem[Condon(1992)]{cond92}
  Condon, J. J. 1992, \araa, 30, 575

\bibitem[Conselice et al.(2000)]{cons00}
  Conselice, C. J., Gallagher, J. S., Calzetti, D., Homeier, N., \& Kinney, 
  A. 2000, \aj, 119, 79

\bibitem[Dallacasa et al.(2000)]{dall00}
  Dallacasa, D., Stanghellini, C., Centonza, M., \& Fanti, R. 2000, \aap, 363, 
  887

\bibitem[De~Mello et al.(2000)]{deme00}
  De~Mello, D. F., Leitherer, C., \& Heckman, T. M. 2000, \apj, 530, 251

\bibitem[Devereux \& Young(1991)]{deve91}
  Devereux, N. A. \& Young, J. S. 1991, \apj, 371, 515

\bibitem[Dickman et al.(1986)]{dick86}
  Dickman, R. L., Snell, R. L., \& Schloerb, F. P. 1986, \apj, 309, 326

\bibitem[Downes et al.(1999)]{down99}
  Downes, D., et al. 1999, \aap, 347, 809
% second author is Neri

\bibitem[Downes \& Solomon(1998)]{down98}
  Downes, D. \& Solomon, P. M. 1998, \apj, 507, 615

\bibitem[Downes \& Solomon(2003)]{down03}
  Downes, D. \& Solomon, P. M. 2003, \apj, 582, 37

\bibitem[Downes et al.(1995)]{down95}
  Downes, D., Solomon, P. M., \& Radford, S. J. E. 1995, \apj, 453, L65

\bibitem[Dumke et al.(2001)]{dumk01}
  Dumke, M., Nieten, C., Thuma, R., Wielebinski, R., \& Walsh, W. 2001, \aap, 
  373, 853

\bibitem[Edge et al.(1999)]{edge99}
  Edge, A. C., Ivison, R. J., Smail, I., Blain, A. W., \& Kneib, J.--P. 1999, 
  \mnras, 306, 599

\bibitem[Ellingson et al.(1996)]{elli96}
  Ellingson, E., Yee, H. K. C., Bechtold, J., \& Elston, R. 1996, \apj, 466, 
  L71

\bibitem[Foucaud et al.(2003)]{fouc03}
  Foucaud, S., McCracken, H. J., Le~F{\` e}vre, O., Arnouts, S., Brodwin, M., 
  Lilly, S. J., Crampton, D., \& Mellier, Y. 2003, \aap, 409, 835

\bibitem[Frayer et al.(1999)]{fray99}
  Frayer, D. T., et al. 1999, \apj, 514, L13

\bibitem[Frayer et al.(1998)]{fray98}
  Frayer, D. T., Ivison, R. J., Scoville, N. Z., Yun, M., Evans, A. S., 
  Smail, I., Blain, A. W., Kneib, J.--P. 1998, \apj, 506, L7

\bibitem[Frayer et al.(1997)]{fray97}
  Frayer, D. T., Papadopoulos, P. P., Bechtold, J., Seaquist, E. R., Yee, 
  H. K. C., \& Scoville, N. Z. 1997, \aj, 113, 562

\bibitem[Giavalisco \& Dickinson(2001)]{giav01}
  Giavalisco, M. \& Dickinson, M. 2001, \apj, 550, 177

\bibitem[Giavalisco et al.(1998)]{giav98}
  Giavalisco, M., Steidel, C. C., Adelberger, K. L., Dickinson, M. E., 
  Pettini, M., \& Kellogg, M. 1998, \apj, 503, 543

\bibitem[Guilloteau \& Lucas(2000)]{guil00}
  Guilloteau, S. \& Lucas, R. 2000, in Imaging at Radio Through Submillimeter 
  Wavelengths, ed. J. G. Mangum \& S. J. E. Radford (San Francisco: ASP), 299

\bibitem[Guilloteau et al.(1992)]{guil92}
  Guilloteau, S., et al. 1992, \aap, 262, 624
% second author is Delannoy

\bibitem[Guilloteau et al.(1999)]{guil99}
  Guilloteau, S., Omont, A., Cox, P., McMahon, R. G., \& Petitjean, P. 1999, 
  \aap, 349, 363

\bibitem[Heckman et al.(2000)]{heck00}
  Heckman, T. M., Lehnert, M. D., Strickland, D. K., \& Armus, L. 2000, \apjs, 
  129, 493

\bibitem[Heckman et al.(1998)]{heck98}
  Heckman, T. M., Robert, C., Leitherer, C., Garnett, D. R., \& van~der~Rydt, 
  F. 1998, \apj, 503, 646

\bibitem[Helfer et al.(2003)]{helf03}
  Helfer, T. T., Thornley, M. D., Regan, M. W., Wong, T., Sheth, K., Vogel, 
  S. N., Blitz, L., \& Bock, D. C.--J. 2003, \apjs, 145, 259

\bibitem[Howarth(1983)]{howa83}
  Howarth, I. D. 1983, \mnras, 203, 301

\bibitem[Ivison et al.(2000)]{ivis00}
  Ivison, R. J., Smail, I., Barger, A. J., Kneib, J.--P., Blain, A. W., 
  Owen, F. N., Kerr, T. H., \& Cowie, L. L. 2000, \mnras, 315, 209

\bibitem[Kassiola \& Kovner(1993)]{kass93}
  Kassiola, A. \& Kovner, I. 1993, \apj, 417, 450

\bibitem[Kauffmann \& Haehnelt(2000)]{kauf00}
  Kauffmann, G. \& Haehnelt, M. 2000, \mnras, 311, 576 (KH00)

\bibitem[Keeton(2001)]{keet01}
  Keeton, C. R. 2001, \apj, submitted (astro--ph/0102340)
% catalogue of mass models is 0102341 but citations should be to its references

\bibitem[Kennicutt(1998)]{kenn98}
  Kennicutt, R. C., Jr. 1998, \apj, 498, 541

\bibitem[Kennicutt et al.(1994)]{kenn94}
  Kennicutt, R. C., Jr., Tamblyn, P., \& Congdon, C. W. 1994, \apj, 435, 22

\bibitem[Kolatt et al.(1999)]{kola99}
  Kolatt, T. S., et al. 1999, \apj, 523, L109
% second author is Bullock

\bibitem[Laurent--Muehleisen et al.(1997)]{laur97}
  Laurent--Muehleisen, S. A., Kollgaard, R. I., Ryan, P. J., Feigelson, E. D., 
  Brinkmann, W., \& Siebert, J., 1997, \aaps, 122, 235

\bibitem[Lehnert \& Heckman(1996)]{lehn96}
  Lehnert, M. D. \& Heckman, T. M. 1996, \apj, 472, 546

\bibitem[Leitherer et al.(2001)]{leit01}
  Leitherer, C., Le{\~ a}o, J. R. S., Heckman, T. M., Lennon, D. J., Pettini, 
  M., \& Robert, C. 2001, \apj, 550, 724

\bibitem[Leitherer et al.(1999)]{leit99}
  Leitherer, C., et al. 1999, \apjs, 123, 3
% second author is Schaerer

\bibitem[Lisenfeld et al.(1996)]{lise96}
  Lisenfeld, U., Hills, R. E., Radford, S. J. E., \& Solomon, P. M. 1996, in 
  Cold Gas at High Redshift, eds. M. N. Bremer \& N. Malcolm (Dordrecht: 
  Kluwer), 55

\bibitem[Lowenthal et al.(1997)]{lowe97}
  Lowenthal, J. D., et al. 1997, \apj, 481, 673
% second author is Koo

\bibitem[Madau et al.(1996)]{mada96}
  Madau, P., Ferguson, H. C., Dickinson, M. E., Giavalisco, M., Steidel, C. C.,
  \& Fruchter, A. 1996, \mnras, 283, 1388

\bibitem[Mauersberger et al.(1999)]{maue99}
  Mauersberger, R., Henkel, C., Walsh, W., \& Schulz, A. 1999, \aap, 341, 256

\bibitem[Meurer et al.(1999)]{meur99}
  Meurer, G. R., Heckman, T. M., \& Calzetti, D. 1999, \apj, 521, 64

\bibitem[Meurer et al.(1997)]{meur97}
  Meurer, G. R., Heckman, T. M., Lehnert, M. D., Leitherer, C., \& Lowenthal, 
  J. 1997, \aj, 114, 54

\bibitem[Mihos \& Hernquist(1994)]{miho94}
  Mihos, J. C. \& Hernquist, L. 1994, \apj, 425, 13

\bibitem[Mo et al.(1999)]{mo99}
  Mo, H. J., Mao, S., \& White, S. D. M. 1999, \mnras, 304, 175

\bibitem[Moriondo et al.(1998)]{mori98}
  Moriondo, G., Giovanardi, C., \& Hunt, L. K. 1998, \aap, 339, 409

\bibitem[Nakanishi et al.(1997)]{naka97}
  Nakanishi, K., Ohta, K., Takeuchi, T. T., Akiyama, M., Yamada, T., \& Shioya,
  Y. 1997, \pasj, 49, 535

\bibitem[Nandra et al.(2002)]{nand02}
  Nandra, K., Mushotzky, R. F., Arnaud, K., Steidel, C. C., Adelberger, K. L.,
  Gardner, J. P., Teplitz, H. I., \& Windhorst, R. A. 2002, \apj, 576, 625

\bibitem[Neri et al.(2003)]{neri03}
  Neri, R., et al. 2003, \apj, 597, L113
% second author is Genzel

\bibitem[Omont et al.(2003)]{omon03}
  Omont, A., Beelen, A., Bertoldi, F., Cox, P., Carilli, C. L., Priddey, R. S.,
  McMahon, R. G., \& Isaak, K. G. 2003, \aap, 398, 857

\bibitem[Ouchi et al.(1999)]{ouch99}
  Ouchi, M., Yamada, T., Kawai, H., \& Ohta, K. 1999, \apj, 517, L19

\bibitem[Papovich et al.(2001)]{papo01}
  Papovich, C., Dickinson, M., \& Ferguson, H. C. 2001, \apj, 559, 620

\bibitem[Peacock et al.(2000)]{peac00}
  Peacock, J. A., et al. 2000, \mnras, 318, 535
% second author is Rowan--Robinson

\bibitem[Pettini et al.(1998)]{pett98}
  Pettini, M., Kellogg, M., Steidel, C. C., Dickinson, M., Adelberger, K. L., 
  \& Giavalisco, M., 1998, \apj, 508, 539

\bibitem[Pettini et al.(2002)]{pett02}
  Pettini, M., Rix, S. A., Steidel, C. C., Adelberger, K. L., Hunt, M. P., \& 
  Shapley, A. E. 2002, \apj, 569, 742

\bibitem[Pettini et al.(2001)]{pett01}
  Pettini, M., Shapley, A. E., Steidel, C. C., Cuby, J.--G., Dickinson, M., 
  Moorwood, A. F. M., Adelberger, K. L., \& Giavalisco, M. 2001, \apj, 554, 981

\bibitem[Pettini et al.(2000)]{pett00}
  Pettini, M., Steidel, C. C., Adelberger, K. L., Dickinson, M., \& 
  Giavalisco, M., 2000, \apj, 528, 96 

\bibitem[Porciani \& Giavalisco(2002)]{porc02}
  Porciani, C. \& Giavalisco, M. 2002, \apj, 565, 24

\bibitem[Rosolowsky et al.(2003)]{roso03}
  Rosolowsky, E., Plambeck, R., Engargiola. G. \& Blitz, L. 2003, \apj, in
  press (astro--ph/0307322)

\bibitem[Salpeter(1955)]{salp55}
  Salpeter, E. E. 1955, \apj, 121, 161

\bibitem[Sawicki(2001)]{sawi01}
  Sawicki, M. 2001, \aj, 121, 2405

\bibitem[Sawicki \& Yee(1998)]{sawi98}
  Sawicki, M. \& Yee, H. K. C. 1998, \aj, 115, 1329

\bibitem[Schmidt(1959)]{schm59}
  Schmidt, M. 1959, \apj, 129, 243

\bibitem[Scoville \& Young(1991)]{scov91}
  Scoville, N. Z. \& Young, J. S. 1991, \araa, 29, 581

\bibitem[Scoville et al.(1997)]{scov97}
  Scoville, N. Z., Yun, M. S., Windhorst, R. A., Keel, W. C., \& Armus, L. 
  1997, \apj, 485, L21

\bibitem[Seibert et al.(2002)]{seib02}
  Seibert, M., Heckman, T. M., \& Meurer, G. R. 2002, \aj, 124, 46

\bibitem[Seitz et al.(1998)]{seit98} 
  Seitz, S., Saglia, R. P., Bender, R., Hopp, U., Belloni, P., \& Ziegler, B. 
  1998, \mnras, 298, 945 (S98)

\bibitem[Shapley et al.(2001)]{shap01}
  Shapley, A. E., Steidel, C. C., Adelberger, K. L., Dickinson, M., Giavalisco,
  M., \& Pettini M. 2001, \apj, 562, 95

\bibitem[Shapley et al.(2003)]{shap03}
  Shapley, A. E., Steidel, C. C., Pettini, M., \& Adelberger, K. L. 2003, \apj,
  588, 65

\bibitem[Shepherd(1997)]{shep97}
  Shepherd, M. C. 1997, in Astronomical Data Analysis Software and Systems VI, 
  eds. G. Hunt \& H. E. Payne (San Francisco: ASP), 77

\bibitem[Solomon \& Barrett(1991)]{solo91}
  Solomon, P. M. \& Barrett, J. W. 1991, in Dynamics of Galaxies and Their 
  Molecular Cloud Distributions, ed. F. Combes \& F. Casoli (Dordrecht: 
  Kluwer), 235

\bibitem[Solomon et al.(1997)]{solo97}
  Solomon, P. M., Downes, D., Radford, S. J. E., \& Barrett, J. W. 1997, \apj, 
  478, 144

\bibitem[Somerville \& Primack(1999)]{some99}
  Somerville, R. S. \& Primack, J. R. 1999, \mnras, 310, 1087

\bibitem[Somerville et al.(2001)]{some01}
  Somerville, R. S., Primack, J. R., \& Faber, S. M. 2001, \mnras, 320, 504
  (SPF01)

\bibitem[Steidel et al.(1999)]{stei99}
  Steidel, C. C., Adelberger, K. L., Giavalisco, M., Dickinson, M., \& 
  Pettini, M. 1999, \apj, 519, 1

\bibitem[Steidel et al.(2003)]{stei03}
  Steidel, C. C., Adelberger, K. L., Shapley, A. E., Pettini, M., Dickinson, 
  M., \& Giavalisco, M. 2003, \apj, 592, 728

\bibitem[Steidel \& Hamilton(1992)]{stei92}
  Steidel, C. C. \& Hamilton, D. 1992, \aj, 104, 941

\bibitem[Strong et al.(1988)]{stro88}
  Strong, A. W., et al. 1988, \aap, 207, 1
% second author is Bloemen

\bibitem[Teplitz et al.(2002)]{tepl02}
  Teplitz, H. I., Malkan, M. A., \& McLean, I. S. 2002, BAAS, 201, 52.04

\bibitem[Teplitz et al.(2000)]{tepl00}
  Teplitz, H. I., et al. 2000, \apj, 533, L65 
% second author is McLean

\bibitem[Towle et al.(1996)]{towl96}
  Towle, J. P., Feldman, P. A., \& Watson, J. K. G. 1996, \apjs, 107, 747

\bibitem[van~der~Werf et al.(2001)]{vand01}
  van~der~Werf, P. P., Knudsen, K. K., Labb{\' e}, I., \& Franx, M. 2001, in 
  Deep millimeter surveys: implications for galaxy formation and evolution, 
  ed. J. D. Lowenthal \& D. H. Hughes (Singapore: World Scientific Publishing),
  103

\bibitem[van~Moorsel, Kemball, \& Greisen(1996)]{vanm96}
  van~Moorsel, G., Kemball, A., \& Greisen, E. 1996, in Astronomical Data 
  Analysis Software and Systems V, ed. G. H. Jacoby \& J. Barnes (San 
  Francisco: ASP), 37

\bibitem[Webb et al.(2003)]{webb03}
  Webb, T. M., et al. 2003, \apj, 582, 6
% second author is Eales

\bibitem[Wechsler et al.(2001)]{wech01}
  Wechsler, R. H., Somerville, R. S., Bullock, J. S., Kolatt, T. S., Primack, 
  J. R., Blumenthal, G. R., \& Dekel, A. 2001, \apj, 554, 85

\bibitem[Wilson(1995)]{wils95}
  Wilson, C. D. 1995, \apj, 448, L97

\bibitem[Wolfire et al.(1993)]{wolf93}
  Wolfire, M. G., Hollenbach, D., \& Tielens, A. G. G. M. 1993, \apj, 402, 195

\bibitem[Wong \& Blitz(2002)]{wong02}
  Wong, T. \& Blitz, L. 2002, \apj, 569, 157

\bibitem[Yee et al.(1996)]{yee96}
  Yee, H. K. C., Ellingson, E., Bechtold, J., Carlberg, R. G., \& Cuillandre, 
  J.--C. 1996, \aj, 111, 1783

\bibitem[Yun \& Carilli(2002)]{yun02}
  Yun, M. S. \& Carilli, C. L. 2002, \apj, 568, 88

\end{thebibliography}
\end{document}